\documentclass[journal ]{new-aiaa}
\usepackage[utf8]{inputenc}
\usepackage{textcomp}
\usepackage{graphicx}
\usepackage{amsmath}
\usepackage[version=4]{mhchem}
\usepackage{siunitx}
\usepackage{longtable,tabularx}
\usepackage{hyperref}
\usepackage{latexsym}
\usepackage{multirow}
\usepackage{makecell}
\usepackage{graphicx}
\usepackage{float}
\usepackage{subfigure}
\usepackage{subcaption}

\title{Single-Impulse Reachable Set in Arbitrary Dynamics Using Polynomials}

\author{Xingyu Zhou\footnote{Ph.D. Candidate, School of Aerospace Engineering; \url{zhouxingyu@bit.edu.cn}.}}
\affil{Beijing Institute of Technology, 100081 Beijing, People’s Republic of China}
\author{Roberto Armellin\footnote{Professor, Te Pūnaha Ātea - Space Institute, 20 Symonds Street, Auckland Central; \url{roberto.armellin@auckland.ac.nz} (Member AIAA).}}
\affil{University of Auckland, Auckland 1010, New Zealand}
\author{Dong Qiao\footnote{Professor, School of Aerospace Engineering; \url{qiaodong@bit.edu.cn} (Corresponding Author).} and Xiangyu Li\footnote{Associate Professor, School of Aerospace Engineering; \url{lixiangy@bit.edu.cn}.}}
\affil{Beijing Institute of Technology, 100081 Beijing, People’s Republic of China}

\begin{document}

\maketitle

\begin{abstract}
This paper presents a method to determine the reachable set (RS) of spacecraft after a single velocity impulse with an arbitrary direction, which is appropriate for the RS in both the state and observation spaces under arbitrary dynamics, extending the applications of current RS methods from two-body to arbitrary dynamics. First, the single-impulse RS model is generalized as a family of two-variable parameterized polynomials in the differential algebra scheme. Then, using the envelope theory, the boundary of RS is identified by solving the envelope equation. A framework is proposed to reduce the complexity of solving the envelope equation by converting it to the problem of searching the root of a one-variable polynomial. Moreover, a high-order local polynomial approximation for the RS envelope is derived to improve computational efficiency. The method successfully determines the RSs of two near-rectilinear halo orbits in the cislunar space. Simulation results show that the RSs in both state and observation spaces can be accurately approximated under the three-body dynamics, with relative errors of less than 0.0658\%. In addition, using the local polynomial approximation, the computational time for solving the envelope equation is reduced by more than 84\%.
\end{abstract}

\section{Introduction}
\lettrine{T}{he} concept of reachable sets (RSs) in orbital mechanics is vital for understanding the dynamics and control of spacecraft maneuvers in diverse gravitational environments \cite{Vinh1995, Xue2010, Holzinger2012, Hall2020CMDA, Jain2023}. An RS encompasses all possible states that a spacecraft can achieve from a given initial state, considering specific control inputs and dynamic constraints over a defined time period \cite{Patel2023JAS}. The RS delineates where the spacecraft can operate effectively. It offers a comprehensive view of the potential outcomes of various maneuvers and has been extensively applied in space-related fields such as debris cloud evolution \cite{Wen2016, Servadio2023, Liu2021ASR}, collision avoidance \cite{Xu2018PG, Wen2022Astro, Pavanello2024}, trajectory optimization \cite{Chen2021JGCD, Cui2024, Bowerfind2024}, maneuver detection \cite{AguilarMarsillach2021, Li2024SST}, and non-cooperative target tracking \cite{Hall2020, Hall2023}.

The RS can be broadly divided into two categories according to the types of maneuvers: the RS with continuous maneuvers and the RS with impulsive maneuvers \cite{Xue2010, Duan2018}. 
In the first category, Holzinger \emph{et al.} introduced a level-set method to explore the time-free RS in terms of orbital elements.
Pang and Wen developed a grid method to determine the time-fixed positional RS while considering finite thrust and mass constraints, wherein the RS boundary is determined by solving optimal control problems on a series of grid points \cite{Pang2022IEEE}. 
Lin and Zhang proposed a distance-fields-over-grids method for solving the continuous-thrust relative RS under energy and fuel constraints \cite{Lin2023RS}.
Wang and Jiang provided an analytical solution to define the envelope of the RS for low-thrust, linearized relative motion \cite{Wang2023}. 
More recently, Natherson and Scheeres proposed a versatile approach to accurately determine the boundary of an RS in a specified direction, incorporating constraints on terminal velocities to ensure safe operations during formation flying, spacecraft rendezvous, and docking scenarios \cite{Natherson2024}.

The second category of RS problems assumes that the spacecraft experiences an instantaneous change in velocity \cite{Xue2010, Wen2014_1, Wen2014_23}; consequently, the evolution of the spacecraft’s orbital state following the velocity impulse is governed by nonlinear dynamics represented by a set of ordinary differential equations (ODEs). Xue \emph{et al.} were the first to investigate the single-impulse RS problem within the framework of ideal two-body dynamics \cite{Xue2010}, focusing on three typical scenarios: 1) velocity impulse at a fixed epoch with an arbitrary direction, 2) velocity impulse at an arbitrary epoch with a fixed direction, and 3) velocity impulse with both arbitrary epoch and arbitrary direction. Their methodology was found to overestimate the size of the RS. Wen \emph{et al.} subsequently enhanced Xue \emph{et al.}’s approach, providing a more accurate envelope of the RS for the first scenario, where the maneuver epoch is fixed, and the impulse direction is arbitrary, by utilizing the terminal velocity hyperbola of the orbital two-point boundary value problem \cite{Wen2014_1}. In Wen \emph{et al.}’s later work, they successfully determined the RS envelope for the remaining two cases by assessing the extreme values of the reachable distance across various directions \cite{Wen2014_23}. 
The above methods can determine the RS for most cases; however, they fail for highly elliptical orbits, whose RS is almost unbounded. Chen \emph{et al.} proposed a maximum reachable distance approach to investigate the single-impulse RS of highly elliptical orbits, which can effectively decide the maximum eccentricity for the nominal trajectory under a maximum available impulse \cite{Chen2018AA}.

Recently, researchers have expanded the application of the notion of RS to various domains, including gravity assist maneuvers, asteroid hopping, and ground track analysis. Chen \emph{et al.} explored the RS of a spacecraft after a gravity-assist flyby, providing a valuable assessment tool to evaluate the flyby reachability for potential targets \cite{Chen2019, Chen2020}. By modeling hyperbolic excess velocities post-flyby through a velocity spherical crown, it was transformed into a single-impulse RS problem with constrained impulse directions instead of the arbitrary impulse direction assumption depicted in previous studies. Building on this work, Cao \emph{et al.} utilized the gravity-assist flyby RS to evaluate multiple gravity-assist opportunities for interplanetary missions \cite{Cao2024}. Wen \emph{et al.} introduced a novel concept of hop RS to assess the surface mobility of a hopping rover for asteroid landing applications \cite{Wen2020Hop}. Zhang \emph{et al.} broadened the scope of the single-impulse RS concept to address ground-tracking problems, focusing on the boundaries of the projections of the spacecraft’s orbit onto the Earth’s surface over time \cite{Zhang2021}.

The state-of-the-art methods on RS are mainly under the ideal unperturbed dynamics as an analytical orbital propagation solution is available in this scenario. They are inapplicable if additional perturbations are considered. Recently, Wen \emph{et al.} proposed a method that can determine the RS under the J2 perturbation \cite{Wen2022JGCD}. After that, the method was applied to model the medium-term evolution of a debris cloud resulting from an on-orbit fragmentation event \cite{Wen2024JGCD}. However, Wen \emph{et al.}’s method still cannot provide an accurate RS when more complex perturbations, for example, the third-body perturbations and higher-order non-spherical perturbations, are considered. These perturbations become particularly important for spacecraft operating in multibody systems \cite{Boone2022, Zhou2022AA, Yang2023} or orbits around small bodies \cite{Li2018JGCD}. In addition, current work mostly focuses on the RS in orbital state space, specifically, the position sub-space, and lacks the investigation of RS in observation space, which is vital for applications such as maneuver detection and sensor tasking.

Motivated by the aforementioned two points, this paper aims to provide a method that can determine the RS in both state and observation spaces under arbitrary dynamics. The proposed method focuses on the first scenario, that is, the RS of a single impulsive maneuver at a fixed epoch with an arbitrary direction. First, the Differential Algebra (DA), combined with the automatic domain split (ADS) technique, is employed to derive high-order polynomials for the target variables with respect to the control parameters of the velocity impulse. As the DA and ADS are capable of handling any dynamics, the proposed method can avoid the requirement of an unperturbed dynamics environment. 
For the RS in state space, two approaches, one based on the partial map inversion technique and the other based on Newton’s iteration, are proposed to project the three-dimensional RS onto an auxiliary plane to reduce the complexity. The partial map inversion approach is faster, whereas Newton’s iteration approach is more robust.
Then, the RS boundary is formulated based on the envelope theory. The envelope equation is approximated using high-order polynomials, whose roots compose a characteristic curve related to the envelope of the RS. A framework is proposed to reduce the complexity of searching roots of the envelope equation. 
Moreover, to improve efficiency, a local polynomial approximation for the RS boundary is derived by implementing a partial map inversion for the high-order polynomials of the envelope equation.
Finally, the proposed method is applied to solve the RS problems in a highly nonlinear dynamic system: cislunar space, with the circular-restricted three-body problem (CRTBP) model adopted as the orbital dynamics.

The remainder of this paper is organized as follows. Section~\ref{Sec:Preliminaries} briefly introduces the DA and ADS. Section~\ref{Sec:Reachable Set Model} derives the high-order polynomials for modeling the two RS problems. The RS boundary is solved via envelope theory in Sec.~\ref{Sec:Methodology for Solving the Reachable Set Boundary}. Simulation examples are presented in Sec.~\ref{Sec:Numerical simulations}, and conclusions are given in Sec.~\ref{Sec:Conclusion}.

\section{Preliminaries} \label{Sec:Preliminaries}

\subsection{Differential Algebra} \label{Sec:Differential Algebra}
\begin{figure}[!h]
    \centering
    \includegraphics[width=0.65\linewidth]{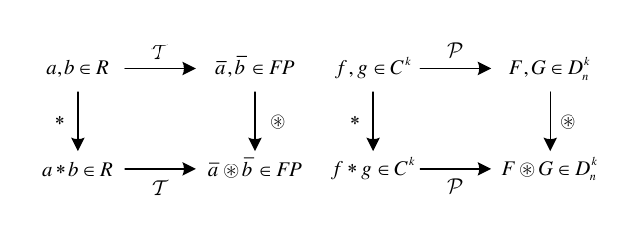}
    \caption{\label{fig:An illustration of two computational frameworks} An illustration of two computational frameworks \cite{Valli2013}.}
\end{figure}

The DA technique was initially developed as an algebraic method for addressing analytical problems \cite{hawkes1999modern}. This work employs the DA to derive high-order Taylor expansions for the flow of nonlinear orbital dynamics with respect to the control parameters of an impulsive maneuver. While traditional numerical algorithms evaluate functions at specific points, the DA takes advantage of the fact that functions can provide more detailed information beyond their numeric values \cite{Armellin2010, Morselli2014, Fu2024Astro}. The core concept of DA involves handling functions and operations on them in a manner analogous to the treatment of real numbers within a computer environment, as illustrated in the diagram in Fig.~\ref{fig:An illustration of two computational frameworks}. The left and right subfigures of Fig.~\ref{fig:An illustration of two computational frameworks} show the floating point (FP) representation of real numbers (in a computer environment) and the algebra of Taylor polynomials (in an DA framework), respectively. The symbol “$\ast$” represents an (arbitrary) operation in real numbers, and the symbol “$\circledast$” is the corresponding adjoint operation for the set of FP numbers. In the left subfigure of Fig.~\ref{fig:An illustration of two computational frameworks}, $\bar{a}$ and $\bar{b}$ denote the FP representations of the two real numbers $a$ and $b$ in the computational environment, respectively, whereas, in the right subfigure, $F$ and $G$ represent the \emph{n}-th order Taylor expansions of the functions $f$ and $g$, respectively. Like the FP arithmetic, one can composite and invert functions, solve nonlinear systems explicitly, treat common elementary functions, and perform differentiation and integration in the DA framework \cite{Valli2013, Feng2019, Fossa2024}.

The Differential Algebra Core Engine (DACE) \footnote{Available at \url{https://github.com/dacelib/dace/}.} is a convenient and powerful DA computational toolbox (C++). The DACEyPy \footnote{Available at \url{https://github.com/giovannipurpura/daceypy/}.} is a Python wrapper of DACE, which is employed in this work for the DA implementation.

\subsection{Automatic Domain Split} \label{Sec:Adaptive Domain Split}

\begin{figure}[!h]
    \centering
    \includegraphics[width=0.5\linewidth]{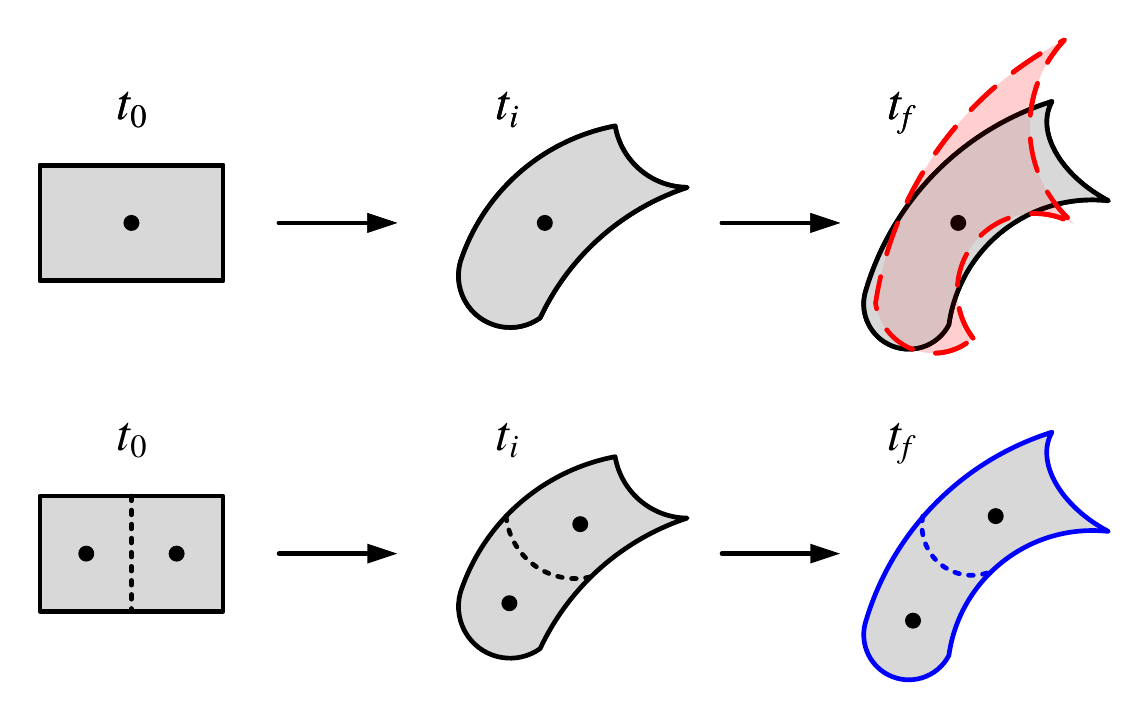}
    \caption{\label{fig:An illustration of ADS} An illustration of ADS \cite{Wittig2015}.}
\end{figure}

The DA has demonstrated its efficiency by substituting thousands of point-wise integrations in Monte Carlo simulations with the rapid computation of arbitrary-order Taylor expansions for the orbital dynamics \cite{Valli2013}. However, the existing DA-based high-order Taylor polynomial method encounters limitations when strong nonlinearities or large uncertainties in the dynamics hinder the Taylor expansion’s convergence in specific directions. To be specific, the control parameters of the single-impulse RS problem are two angles (\emph{i.e.}, the elevation $\alpha$ and azimuth $\beta$) that define the direction of an impulsive maneuver. As the direction of the impulse is arbitrary, the two angles have wide ranges (\emph{i.e.}, $\alpha \in \left [ -\frac{\pi}{2},\frac{\pi}{2} \right ]$ and $\beta \in \left [ -\pi,\pi \right ] $). In this case, a single Taylor polynomial is inadequate to meet the accuracy requirements (as shown in the upper sub-figure in Fig.~\ref{fig:An illustration of ADS}).

To address this challenge, the ADS technique was developed \cite{Wittig2015}. The ADS assumes that the absolute values of the polynomial coefficients decay exponentially as the order increases. Thus, given an \emph{n}-th order Taylor polynomial, the ADS can predict the coefficient size of the $n+1$ order using an exponential fit, which is further employed to evaluate the polynomial’s truncation errors. As shown in Fig.~\ref{fig:An illustration of ADS}, the ADS automatically splits the polynomial expansion of a given initial domain into two separate polynomials (with two subdomains) whenever the truncation error exceeds a predefined threshold $\epsilon$ during propagation. This integrated approach facilitates more accurate predictions of the spacecraft’s behavior. The final results of the ADS are a collection of localized Taylor polynomials, with each polynomial corresponding to an automatically determined subset of the initial domain.

\section{Reachable Set Model} 
\label{Sec:Reachable Set Model}
Consider a general form of nonlinear orbital dynamics governed by the set of ODEs 
\begin{equation} \label{eq:dxdt}
    \frac{{d\boldsymbol{x}}}{{dt}} = \boldsymbol{f}(\boldsymbol{x}) \,,
\end{equation}
where $\boldsymbol{f}:{\mathbb{R}^6} \mapsto {\mathbb{R}^6}$ represents the nonlinear dynamics model, and $\boldsymbol{x} = [\boldsymbol{r};\boldsymbol{v}] \in {\mathbb{R}^6}$ denotes the time-depended orbital state, with $\boldsymbol{r} \in {\mathbb{R}^3}$ and $\boldsymbol{v} \in {\mathbb{R}^3}$ being position and velocity vectors, respectively. Given an orbital state vector $\boldsymbol{x}_{0} = [\boldsymbol{r}_{0};\boldsymbol{v}_{0}] \in {\mathbb{R}^6}$ at the initial epoch $t_0$, one can determine the orbital state at any future epoch $t$ by integrating the ODE in Eq.~\eqref{eq:dxdt} as
\begin{equation} \label{eq:xt}
    \boldsymbol{x} = \int\limits_{\tau  = 0}^t {\boldsymbol{f}(\boldsymbol{x})d\tau }  + {\boldsymbol{x}_0} \,.
\end{equation}

Assume that at the initial epoch $t_0$, the spacecraft executes an impulsive maneuver $\Delta {\boldsymbol{v}_0}$, expressed as
\begin{equation} \label{eq:dv}
    \Delta {\boldsymbol{v}_0} = {[\Delta {v_{x,0}},\Delta {v_{y,0}},\Delta {v_{z,0}}]^T} \in {\mathbb{R}^3} \,,
\end{equation}
where $\Delta {v_{x,0}}$, $\Delta {v_{y,0}}$, and $\Delta {v_{z,0}}$ denote the velocity impulse along the \emph{x}-, \emph{y}-, and \emph{z} axes, respectively. Then, using the spherical coordinate, the mathematical model of the velocity impulse is rewritten as
\begin{equation} \label{eq:dv_sph}
    \begin{aligned}
        & \Delta {v_{x,0}} = \left\| {\Delta {\boldsymbol{v}_0}} \right\| \cdot \cos \alpha \cos \beta \\
        & \Delta {v_{y,0}} = \left\| {\Delta {\boldsymbol{v}_0}} \right\| \cdot \cos \alpha \sin \beta \\
        & \Delta {v_{z,0}} = \left\| {\Delta {\boldsymbol{v}_0}} \right\| \cdot \sin \alpha 
    \end{aligned} \,,
\end{equation}
where $\left\| {\Delta {\boldsymbol{v}_0}} \right\|$ represents the magnitude of the velocity impulse, which is upper bounded by a maximum velocity increment $\Delta {v_{\max }}$, and $\alpha$ and $\beta$ are two control parameters (\emph{i.e.}, elevation and azimuth) that characterize the direction of the velocity impulse, respectively.

According to Ref.~\cite{Wen2022JGCD}, the boundary of the RS is determined by the maximal velocity impulse, that is, $\left\| {\Delta {\boldsymbol{v}_0}} \right\| \equiv \Delta {v_{\max }}$. In this case, Eq.~\eqref{eq:dv_sph} can be rewritten as
\begin{equation} \label{eq:dvmax_sph}
    \begin{aligned}
        & \Delta {v_{x,0}} = \Delta {v_{\max }} \cdot \cos \alpha \cos \beta \\
        & \Delta {v_{y,0}} = \Delta {v_{\max }} \cdot \cos \alpha \sin \beta \\
        & \Delta {v_{z,0}} = \Delta {v_{\max }} \cdot \sin \alpha 
    \end{aligned} \,,
\end{equation}
which, in the DA framework, can be further formulated using polynomials as
\begin{equation} \label{eq:dv_polynomial}
    \Delta {\boldsymbol{v}_0} \approx {{\cal T}_{\Delta {\boldsymbol{v}_0}}}(\alpha ,\beta ) \,.
\end{equation}

Then, based on the integration formulation in Eq.~\eqref{eq:dv_polynomial}, one can derive the high-order Taylor expansions for the orbital state $\boldsymbol{x}$ with respect to the two control parameters $\alpha$ and $\beta$ in the DA scheme as
\begin{equation} \label{eq:x_polynomial}
    \boldsymbol{x} \approx {\cal T}_{\boldsymbol{x}}(\alpha ,\beta ) \,.
\end{equation}
In this work, the DA’s integration is implemented using an $8^{\rm th}$-order Runge-Kutta (RK78) scheme with a relative tolerance of $10^{-12}$ and an absolute tolerance of $10^{-12}$. In the following two subsections, the mathematical models for the RSs in state and observation spaces will be discussed using the high-order Tyalor polynomials in the DA framework.

\subsection{Reachable Set in State Space} 
\label{Sec:Reachable Set in State Space}
\begin{figure}[!h]
    \centering
    \includegraphics[width=0.7\linewidth]{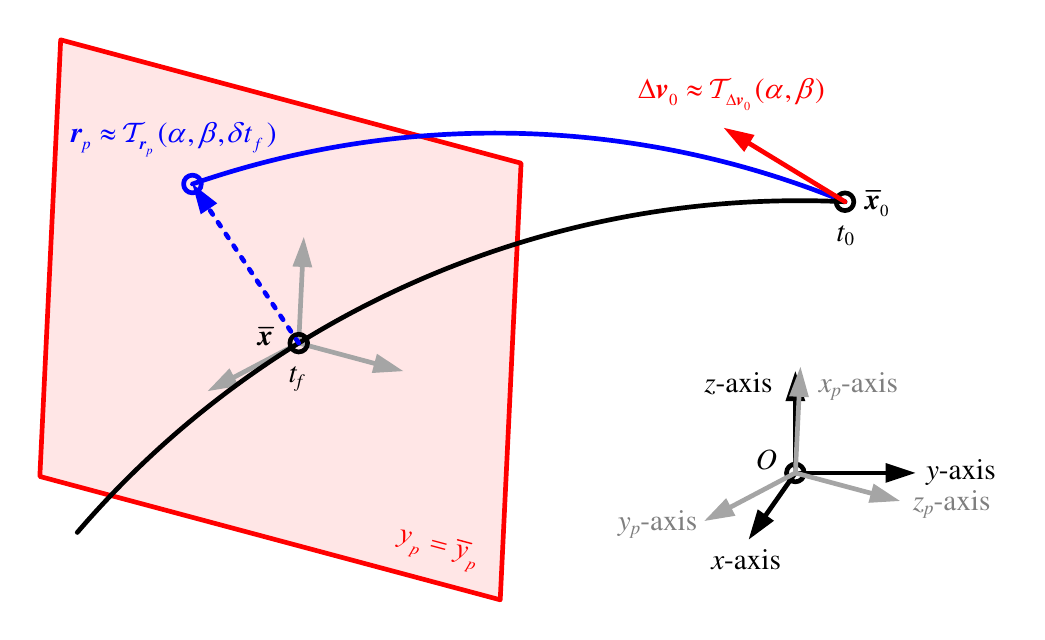}
    \caption{\label{fig:An illustration of the characterized plane} An illustration of the auxiliary plane.}
\end{figure}

In previous works, the RS in state space usually refers to position RS, whose boundary can be described using a three-dimensional surface or a collection of three-dimensional points. In this work, an auxiliary plane is defined to reduce the complexity of determining the RS in state space. By projecting the three-dimensional RS onto this auxiliary plane, one only needs to determine the RS’s boundary in a two-dimensional subspace. As graphically illustrated in Fig.~\ref{fig:An illustration of the characterized plane}, the auxiliary plane is selected to be orthogonal to the velocity direction of the non-maneuver nominal trajectory. Let ${\boldsymbol{\bar x}_0} = [{\boldsymbol{\bar r}_0};{\boldsymbol{\bar v}_0}] \in {\mathbb{R}^6}$ be the initial state of the spacecraft without executing any impulsive maneuver. By propagating the initial state ${\boldsymbol{\bar x}_0}$ under the dynamics in Eq.~\eqref{eq:dxdt}, one can obtain the (nominal) orbital state at a given epoch $t$, labeled as $\boldsymbol{\bar x} = [\boldsymbol{\bar r};\boldsymbol{\bar v}] \in {\mathbb{R}^6}$. A transformation matrix $\boldsymbol{T}(\boldsymbol{\bar x}) \in {\mathbb{R}^{3 \times 3}}$ is then defined as
\begin{equation} \label{eq:Tx}
    \boldsymbol{T}(\boldsymbol{\bar x}) = {\left[ {\frac{{\boldsymbol{\bar h} \times \boldsymbol{\bar v}}}{{\left\| {\boldsymbol{\bar h} \times \boldsymbol{\bar v}} \right\|}},\frac{{\boldsymbol{\bar v}}}{{\left\| {\boldsymbol{\bar v}} \right\|}},\frac{{\boldsymbol{\bar h}}}{{\left\| {\boldsymbol{\bar h}} \right\|}}} \right]^T} \,,
\end{equation}
where $\boldsymbol{\bar h} = \boldsymbol{\bar r} \times \boldsymbol{\bar v}$ represents the angular momentum vector at the epoch $t$.

Given any position vector $\boldsymbol{r} = [x,y,z]^{T} \in {\mathbb{R}^3}$, a new position vector is obtained by performing a coordinate transformation using the matrix $\boldsymbol{T}(\boldsymbol{\bar x})$ as
\begin{equation} \label{eq:rp}
    \boldsymbol{r}_p = {[{x_p},{y_p},{z_p}]^T}\boldsymbol{ = T}(\boldsymbol{\bar x})\boldsymbol{r} \in {\mathbb{R}^3} \,.
\end{equation}
Here, the variable $y_p$ represents the coordinate along the velocity direction. Similarly, for the non-maneuver nominal trajectory, one has
\begin{equation} \label{eq:rp_nomial}
    \boldsymbol{\bar r}_p = {[{\bar x_p},{\bar y_p},{\bar z_p}]^T}\boldsymbol{ = T}(\boldsymbol{\bar x})\boldsymbol{\bar r} \in {\mathbb{R}^3} \,.
\end{equation}
Then, the mathematical model of the selected auxiliary plane is formulated as
\begin{equation} \label{eq:plane}
    \delta {y_p} = {y_p} - {\bar y_p} = 0 \,.
\end{equation}
Therefore, the RS projected on the auxiliary plane can be parameterized using the two variables $x_p$ and $z_p$, and we need to derive the map from the control parameters ($\alpha$ and $\beta$) to the two variables $x_p$ and $z_p$. Here, we propose two ways to derive such a map under the DA framework. The first is a partial map inversion approach, whereas the second one is based on the Newton's iteration. The partial map inversion approach is efficient as it only needs one integration; however, it suffers from a potential numerical singularity when performing polynomial inversion, especially when coming across highly nonlinear dynamics (\emph{e.g.}, at a perilune of an NRHO in the Earth-Moon system). The Newton's iteration approach can avoid inversion operations but is more computationally expensive.

\subsubsection{Partial Map Inversion Approach} \label{sec:Partial Map Inversion Approach}

Substituting Eqs.~\eqref{eq:Tx}-\eqref{eq:rp} into Eq.~\eqref{eq:x_polynomial}, one can construct a high-order Taylor polynomial for the transformed position vector $\boldsymbol{r}_p$ as
\begin{equation} \label{eq:rp_polynomial}
    {\boldsymbol{r}_p} \approx {{\cal T}_{{\boldsymbol{r}_p}}}(\alpha ,\beta ) \,.
\end{equation}
As a velocity impulse $\Delta \boldsymbol{v}_{0}$ is added at the initial epoch, the maneuvered trajectory will deviate from the non-maneuver nominal one; consequently, the transformed position vector $\boldsymbol{r}_p$ may be located outside the selected auxiliary plane. To find the projection of the maneuvered flow, a variation $\delta t_f$ is added to the final epoch $t_f$; thus, Eq.~\eqref{eq:rp_polynomial} can be rewritten as
\begin{equation} \label{eq:rp_t_polynomial}
    {\boldsymbol{r}_p} \approx {{\cal T}_{{\boldsymbol{r}_p}}}(\alpha ,\beta ,\delta {t_f}) \,,
\end{equation}
which, after being substituted into Eq.~\eqref{eq:plane}, yields
\begin{equation} \label{eq:dyp_polynomial}
    \delta {y_p} = {y_p} - {\bar y_p} \approx {{\cal T}_{\delta {y_p}}}(\alpha ,\beta ,\delta {t_f}) \,.
\end{equation}

The high-order Taylor polynomial in Eq.~\eqref{eq:rp_t_polynomial} can be obtained by substituting the independent time variable $t$ in the dynamics model (\emph{i.e.}, Eq.~\eqref{eq:dxdt}) with $\tau  = \frac{t-t_0}{t_f} \in [0,1]$, adding the final epoch $t_f$ as a DA variable and integrating the ODE
\begin{equation} \label{eq:time_expand_ODE}
    \frac{{d\boldsymbol{x}}}{{d\tau }} = \boldsymbol{f}(\boldsymbol{x}) \cdot ({t_f} - {t_0}),\,\tau ({t_0}) = 0
\end{equation}
from $\tau=0$ to $\tau=1$ in the DA scheme \cite{Fu2024Astro}.

Substituting Eq.~\eqref{eq:dyp_polynomial} into Eq.~\eqref{eq:plane}, applying the plane constraint, and using the partial map inversion technique \cite{Valli2013}, one can derive the high-order Taylor polynomial for $\delta t_f$ with respect to the two control parameters $\alpha$ and $\beta$. The steps are given as follows. First, by introducing two identified maps, $\alpha = {{\cal I}_\alpha}(\alpha)$ and $\beta = {{\cal I}_\beta}(\beta)$, the map in Eq.~\eqref{eq:dyp_polynomial} is augmented as
\begin{equation} \label{eq:dyp_augmented_map}
    \left[ {\begin{array}{*{20}{c}}
        \alpha \\
        \beta \\
        {\delta {y_p}}
        \end{array}} \right] \approx \left[ {\begin{array}{*{20}{c}}
            {{{\cal I}_\alpha }}\\
            {{{\cal I}_\beta }}\\
            {{{\cal T}_{\delta {y_p}}}}
        \end{array}} \right] \cdot \left[ {\begin{array}{*{20}{c}}
        \alpha \\
        \beta \\
        {\delta {t_f}}
    \end{array}} \right] \,.
\end{equation}
After performing an inversion for the augmented map in Eq.~\eqref{eq:dyp_augmented_map}, one has
\begin{equation} \label{eq:dyp_augmented_map_invert}
    \left[ {\begin{array}{*{20}{c}}
        \alpha \\
        \beta \\
        {\delta {t_f}}
        \end{array}} \right] \approx {\left[ {\begin{array}{*{20}{c}}
            {{{\cal I}_\alpha }}\\
            {{{\cal I}_\beta }}\\
            {{{\cal T}_{\delta {y_p}}}}
        \end{array}} \right]^{ - 1}} \cdot \left[ {\begin{array}{*{20}{c}}
        \alpha \\
        \beta \\
        {\delta {y_p}}
    \end{array}} \right] \,.
\end{equation}

Applying the constraint of the auxiliary plane to enforce $\delta y_p=0$, the third row of the map in Eq.~\eqref{eq:dyp_augmented_map_invert} becomes
\begin{equation} \label{eq:dtf_polynomial}
    \delta {t_f} \approx {{\cal T}_{\delta {t_f}}}(\alpha ,\beta ,\delta {y_p} = 0) = {{\cal T}_{\delta {t_f}}}(\alpha ,\beta ) \,,
\end{equation}
which delivers how the control parameters $\alpha$ and $\beta$ affect the time variation $\delta {t_f}$ to reach the auxiliary plane. Finally, substituting Eq.~\eqref{eq:dtf_polynomial} back into Eq.~\eqref{eq:rp_t_polynomial}, and removing the $\delta {t_f}$ degree of freedom, one has
\begin{equation} \label{eq:xp_polynomial}
    {x_p} \approx {{\cal T}_{{x_p}}}(\alpha ,\beta ) \,,
\end{equation}
\begin{equation} \label{eq:zp_polynomial}
    {z_p} \approx {{\cal T}_{{z_p}}}(\alpha ,\beta ) \,.
\end{equation}
Equations~\eqref{eq:xp_polynomial}-\eqref{eq:zp_polynomial} are the polynomial approximations for $x_p$ and $z_p$, which correspond to a two-dimensional curve parameterized by the two control variables $\alpha$ and $\beta$. Then, the RS projected on the selected auxiliary plane can be expressed as a family of all possible $x_p$-$z_p$ pairs with the elevation $\alpha$ and azimuth $\beta$ as control parameters. To resolve the RS on the selected plane, one needs to determine the boundary of a family of two-dimensional curves controlled by the parameters $\alpha$ and $\beta$ (\emph{i.e.}, Eqs.~\eqref{eq:xp_polynomial}-\eqref{eq:zp_polynomial}).

\subsubsection{Newton's Iteration Approach} \label{sec:Newton's Iteration Approach}
One can see from Eqs.~\eqref{eq:rp_polynomial}-\eqref{eq:zp_polynomial} that only one integration (\emph{i.e.}, the integration of Eq.~\eqref{eq:time_expand_ODE} to obtain Eq.~\eqref{eq:dyp_polynomial}) is required for the partial map inversion approach. However, the map inversion operation in Eq.~\eqref{eq:dyp_augmented_map_invert} may suffer from a numerical singularity, leading to an inaccurate estimation of the RS (the singularity problem will be numerically shown in Sec.~\ref{Sec:Reachable Set in State Space: 9:2 NRHO Case}). To resolve this problem, an alternative way, noted as Newton’s iteration approach, is presented to generate the polynomials in Eqs.~\eqref{eq:xp_polynomial}-\eqref{eq:zp_polynomial} herein.

First, implement an integration with the time span $t_f$ as the only DA variable to obtain
\begin{equation} \label{eq:yp_tf_polynomial}
    \delta {y_p} \approx {{\cal T}_{\delta {y_p}}}(\delta {t_f}) \,,
\end{equation}
which is a one-variable polynomial with a constant term equal to zero. Then, using Eq.~\eqref{eq:yp_tf_polynomial}, one can compute the derivative of $\delta y_p$ with respect to $t_f$ as $A = \frac{{\partial \delta {y_p}}}{{\partial \delta {t_f}}}$.

Next, let $k=0$ and integrate the trajectory with both $\alpha$ and $\beta$ as DA variables from $t_0$ to $t_f^{(k)} = {t_f}$ to obtain
\begin{equation} \label{eq:yp_angles_polynomial}
    \delta y_p^{(k)} \approx {{\cal T}_{\delta y_p^{(k)}}}(\alpha ,\beta ) \,.
\end{equation}
According to Eq.~\eqref{eq:yp_tf_polynomial}, we have
\begin{equation} \label{eq:yp_A}
    \delta y_p^{(k + 1)} - \delta y_p^{(k)} \approx \frac{{\partial \delta {y_p}}}{{\partial \delta {t_f}}}(t_f^{(k + 1)} - t_f^{(k)}) = A \cdot (t_f^{(k + 1)} - t_f^{(k)}) \,,
\end{equation}
which indicates that we can correct the final epoch as
\begin{equation} \label{eq:tf_correction}
    t_f^{(k + 1)} = t_f^{(k)} - \frac{1}{A}\delta y_p^{(k)} \approx t_f^{(k)} - \frac{1}{A}{{\cal T}_{\delta y_p^{(k)}}}(\alpha ,\beta ) \,.
\end{equation}
Note that in Eq.~\eqref{eq:tf_correction}, the corrected epoch $t_f^{(k + 1)}$ is a polynomial with $\alpha$ and $\beta$ as inputs. Finally, one should iteratively correct the final epoch $t_f^{(\ast)}$ until a pre-defined threshold $\eta$ is met: $\left| {\delta y_p^{(\ast)}} \right| \le \eta $. After determining the final epoch $t_f^{(\ast)}$ when the maneuvered trajectory can reach the auxiliary plane, one can easily obtain the polynomials of $x_p$ and $z_p$ by integrating from $t_0$ to $t_f^{(\ast)}$.

The advantage of Newton’s iteration method lies in it avoids the inversion operation. Its advantage over the partial map inversion approach will be shown in Sec.~\ref{Sec:Reachable Set in State Space: 9:2 NRHO Case}, with a highly nonlinear perilune of a 9:2 NRHO as the example. However, it requires multiple integrations, resulting in a larger computational burden. In practice, the numerical inaccuracies of the map inversion only occurs in extremely nonlinear points. Thus, one can first use the map inversion approach and then detect if that issue occurs. A simple check is provided herein. Substituting the inverted map in Eq.~\eqref{eq:dyp_augmented_map_invert} into the direct map in Eq.~\eqref{eq:dyp_augmented_map}, ideally, one should get an identity map (\emph{i.e.}, the second- and higher-order coefficients are zero). If a coefficient larger than a given pre-defined threshold ($10^{-3}$ in this work) is found, one can use the Newton iteration approach as a replacement.

The method for determining the boundary of the two-dimensional RS will be given in Sec.~\ref{Sec:Methodology for Solving the Reachable Set Boundary}. After computing the boundary of the RS on the plane, the full envelope surface of the three-dimensional position RS can be obtained by simply moving the auxiliary plane along the nominal trajectory. As an example, in the simulation section (\emph{i.e.}, Sec.~\ref{Sec:Numerical simulations}), the three-dimensional RS of a near-rectilinear halo orbit (NRHO) in the cislunar space will be shown in Fig.~\ref{fig:Envelopes of the state RSs at different epochs}.

\subsection{Reachable Set in Observation Space} 
\label{Sec:Reachable Set in Observation Space}
Instead of using an auxiliary plane here we project the RS in the observation space. In particular, an angle-only observation from a single observer is considered. Assume that an observer can measure the line-of-sight (LOS) direction from the observer to the target. Let ${\boldsymbol{r}_O} = {[{x_O},{y_O},{z_O}]^T} \in {\mathbb{R}^3}$ be the position vector of the observer at the given epoch $t$, in the same coordinate system as the target. The LOS observation can be described using two angles as
\begin{equation} \label{eq:gamma}
    \gamma  = {\tan ^{ - 1}}\frac{{x - {x_O}}}{{y - {y_O}}} \,,
\end{equation}
\begin{equation} \label{eq:delta}
    \delta  = {\sin ^{ - 1}}\frac{{x - {x_O}}}{{\left\| {\boldsymbol{r} - {\boldsymbol{r}_O}} \right\|}} \,,
\end{equation}
where $\gamma$ and $\delta$ denote the azimuth and elevation of the angle observation, respectively.

Under the DA framework, polynomial maps can be derived for the azimuth $\gamma$ and elevation $\delta$ as
\begin{equation} \label{eq:gamma_polynomial}
    \gamma  \approx {{\cal T}_\gamma }(\alpha ,\beta ) \,,
\end{equation}
\begin{equation} \label{eq:delta_polynomial}
    \delta  \approx {{\cal T}_\delta }(\alpha ,\beta ) \,.
\end{equation}
Equations~\eqref{eq:gamma_polynomial}-\eqref{eq:delta_polynomial} also represent a family of two-dimensional curves controlled by the parameters $\alpha$ and $\beta$.

\section{Methodology for Solving the Reachable Set Boundary} 
\label{Sec:Methodology for Solving the Reachable Set Boundary}
This section presents the methodology for determining the RS boundary in both state and observation spaces. Section~\ref{Sec:Framework for Solving the Envelope Equation} first presents a framework for identifying the RS boundary by solving the envelope equation. Then, an alternative way is proposed in Sec.~\ref{Sec:Local Polynomial Approximation of the Envelope} to predict the envelope points for the RS, which can significantly reduce the computational cost of solving envelope equations. Finally, the overall procedure for solving the RS boundary is given in Sec.~\ref{Sec:Overall Procedure}.

\subsection{Framework for Solving the Envelope Equation} 
\label{Sec:Framework for Solving the Envelope Equation}
As discussed in Sec.~\ref{Sec:Reachable Set in State Space} and Sec.~\ref{Sec:Reachable Set in Observation Space}, the nonlinear maps of the RSs in both state and observation spaces can be expressed as a collection of two-variable parameterized polynomials. As these two cases share the same mathematical format (\emph{i.e.}, the two-dimensional curves in Eqs.~\eqref{eq:xp_polynomial}-\eqref{eq:zp_polynomial} and Eqs.~\eqref{eq:gamma_polynomial}-\eqref{eq:delta_polynomial}), the RS in state space is taken as an example to show the method. 

To begin with, the ADS technique is employed to split the initial domain of the control parameters, 
\begin{equation} \label{eq:initial_domain}
    {\cal D} = \left\{ {\left. {(\alpha ,\beta )} \right|\alpha \in \left [-\frac{\pi}{2}, \frac{\pi}{2} \right ],\beta  \in \left [ - \pi ,\pi \right ]} \right\} \,.
\end{equation}
yielding $N$ sub-domains as
\begin{equation} \label{eq:sub_domain}
    {\cal D} = {{\cal D}_1} \cup {{\cal D}_1} \cup  \cdots  \cup {{\cal D}_N} \,.
\end{equation}
In Eq.~\eqref{eq:sub_domain}, the \emph{i}-th sub-domain, ${{\cal D}_i}$, can be expressed as
\begin{equation} \label{eq:sub_domain_i}
    {{\cal D}_i} = \left\{ {\left. {(\alpha ,\beta )} \right|\alpha  = {\alpha _i} + {{\cal A}_i} \delta \alpha_{i} ,\beta  = {\beta _i} + {{\cal B}_i} \delta \beta_{i}}  \right\} \,,
\end{equation}
where $\delta \alpha_{i} \in [-1,1]$ and $\delta \beta_{i} \in [-1,1]$ are two DA variables and ${\cal A}_i$ and ${\cal B}_i$ are the corresponding coefficients. The coefficients ${\cal A}_i$ and ${\cal B}_i$ are automatically determined by the ADS. Substituting Eq.~\eqref{eq:sub_domain_i} into Eqs.~\eqref{eq:xp_polynomial}-\eqref{eq:zp_polynomial} generates the polynomials for the \emph{i}-th sub-domain, given as
\begin{equation} \label{eq:xp_delta_polynomial}
    {x_{p,i}} \approx {{\cal T}_{{x_{p,i}}}}(\delta \alpha_{i} ,\delta \beta_{i} ) \,,
\end{equation}
\begin{equation} \label{eq:zp_delta_polynomial}
    {z_{p,i}} \approx {{\cal T}_{{z_{p,i}}}}(\delta \alpha_{i} ,\delta \beta_{i} ) \,.
\end{equation}
Note that in the following of this paper, the subscript $i$ in Eqs.~\eqref{eq:xp_delta_polynomial}-\eqref{eq:zp_delta_polynomial} is omitted for a lighter notation. Therefore, one has, ${x_{p}} \approx {{\cal T}_{{x_{p}}}}(\delta \alpha ,\delta \beta)$ and ${z_{p}} \approx {{\cal T}_{{z_{p}}}}(\delta \alpha ,\delta \beta)$.

The boundary of the RS is determined by identifying the envelopes of the two-variable parameterized polynomials (\emph{i.e.}, Eqs.~\eqref{eq:xp_delta_polynomial}-\eqref{eq:zp_delta_polynomial}). According to the envelope theory \cite{Hohn1988Boot}, the envelope points should satisfy the following condition (known as the envelope equation):
\begin{equation} \label{eq:g}
    g(\delta \alpha ,\delta \beta ) = \frac{{\partial {x_p}}}{{\partial \delta \alpha }}\frac{{\partial {z_p}}}{{\partial \delta \beta }} - \frac{{\partial {x_p}}}{{\partial \delta \beta }}\frac{{\partial {z_p}}}{{\partial \delta \alpha }} = 0 \,.
\end{equation}
Write Eq.~\eqref{eq:xp_delta_polynomial} in a different form as
\begin{equation} \label{eq:xp_polynomial_expand}
    {x_p} \approx {{\cal T}_{{x_p}}}(\delta \alpha ,\delta \beta ) = \sum\limits_{{p_1} + {p_2} \le N} {c_{{p_1}{p_2}}^{{x_p}} \cdot \delta {\alpha ^{{p_1}}}\delta {\beta ^{{p_2}}}} \,,
\end{equation}
where $c_{{p_1}{p_2}}^{{x_p}}$ is the coefficient of the polynomial ${{\cal T}_{{x_p}}}(\delta \alpha ,\delta \beta )$. Using Eq.~\eqref{eq:xp_polynomial_expand}, one can derive
\begin{equation} \label{eq:dxp_alpha}
    \frac{{\partial {x_p}}}{{\partial \delta \alpha }} \approx {{\cal T}_{{{\partial {x_p}}}/{{\partial \delta \alpha }}}}(\delta \alpha ,\delta \beta ) = \sum\limits_{{p_1} + {p_2} \le n} {{p_1}c_{{p_1}{p_2}}^{{x_p}} \cdot \delta {\alpha ^{{p_1} - 1}}\delta {\beta ^{{p_2}}}} \,,
\end{equation}
\begin{equation} \label{eq:dxp_beta}
    \frac{{\partial {x_p}}}{{\partial \delta \beta }} \approx {{\cal T}_{{{\partial {x_p}}}/{{\partial \delta \beta }}}}(\delta \alpha ,\delta \beta ) = \sum\limits_{{p_1} + {p_2} \le n} {{p_2}c_{{p_1}{p_2}}^{{x_p}} \cdot \delta {\alpha ^{{p_1}}}\delta {\beta ^{{p_2} - 1}}} \,.
\end{equation}

In the same way, one finds
\begin{equation} \label{eq:dzp_alpha}
    \frac{{\partial {z_p}}}{{\partial \delta \alpha }} \approx {{\cal T}_{{{\partial {z_p}}}/{{\partial \delta \alpha }}}}(\delta \alpha ,\delta \beta ) = \sum\limits_{{p_1} + {p_2} \le n} {{p_1}c_{{p_1}{p_2}}^{{z_p}} \cdot \delta {\alpha ^{{p_1} - 1}}\delta {\beta ^{{p_2}}}} \,,
\end{equation}
\begin{equation} \label{eq:dzp_beta}
    \frac{{\partial {z_p}}}{{\partial \delta \beta }} \approx {{\cal T}_{{{\partial {z_p}}}/{{\partial \delta \beta }}}}(\delta \alpha ,\delta \beta ) = \sum\limits_{{p_1} + {p_2} \le n} {{p_2}c_{{p_1}{p_2}}^{{z_p}} \cdot \delta {\alpha ^{{p_1}}}\delta {\beta ^{{p_2} - 1}}} \,.
\end{equation}

Substitute Eqs.~\eqref{eq:dxp_alpha}-\eqref{eq:dzp_beta} into Eq.\eqref{eq:g}, and then the envelope equation can be approximated using high-order Taylor polynomials as
\begin{equation} \label{eq:g_polynomial}
    \begin{aligned}
        g(\delta \alpha ,\delta \beta ) &= \frac{{\partial {x_p}}}{{\partial \delta \alpha }}\frac{{\partial {z_p}}}{{\partial \delta \beta }} - \frac{{\partial {x_p}}}{{\partial \delta \beta }}\frac{{\partial {z_p}}}{{\partial \delta \alpha }} \approx {{\cal T}_g}(\delta \alpha ,\delta \beta )\\
         &= \left( {\sum\limits_{{p_1} + {p_2} \le n} {{p_1}c_{{p_1}{p_2}}^{{x_p}} \cdot \delta {\alpha ^{{p_1} - 1}}\delta {\beta ^{{p_2}}}} } \right) \times \left( {\sum\limits_{{p_1} + {p_2} \le n} {{p_2}c_{{p_1}{p_2}}^{{z_p}} \cdot \delta {\alpha ^{{p_1}}}\delta {\beta ^{{p_2} - 1}}} } \right)\\
         &- \left( {\sum\limits_{{p_1} + {p_2} \le n} {{p_2}c_{{p_1}{p_2}}^{{x_p}} \cdot \delta {\alpha ^{{p_1}}}\delta {\beta ^{{p_2} - 1}}} } \right) \times \left( {\sum\limits_{{p_1} + {p_2} \le n} {{p_1}c_{{p_1}{p_2}}^{{z_p}} \cdot \delta {\alpha ^{{p_1} - 1}}\delta {\beta ^{{p_2}}}} } \right)
    \end{aligned} \,.
\end{equation}
Thus, the goal has been changed to find the roots of a two-variable parameterized polynomial equation
\begin{equation} \label{eq:g_constraint}
    g(\delta \alpha ,\delta \beta ) \approx {{\cal T}_g}(\delta \alpha,\delta \beta ) = 0 
\end{equation}
in the search space ${\tilde {\cal D}_i} = \left\{ {\left. {(\delta \alpha ,\delta \beta )} \right| - 1 \le \delta \alpha ,\delta \beta  \le 1} \right\}$. Solving that polynomial equation is not easy because it has two variables and may have zero root, one root, or multiple roots in the given region. A framework is then proposed to search for the polynomial’s roots herein.

\begin{figure}[!h]
    \centering
    \includegraphics[width=0.7\linewidth]{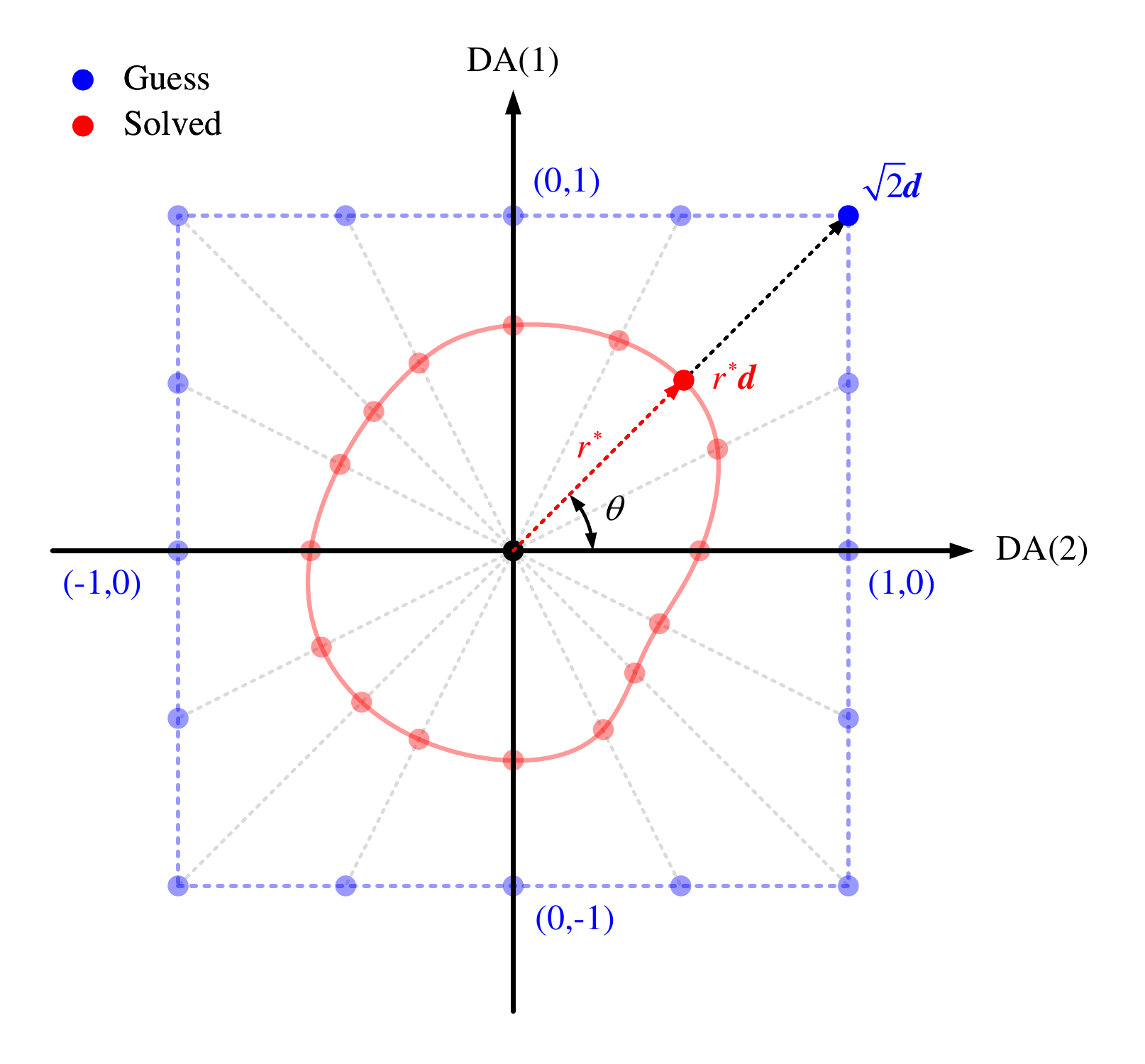}
    \caption{\label{fig:Framework for solving the polynomial of the envelope equation} Framework for solving the polynomial of the envelope equation.}
\end{figure}

As shown in Fig.~\ref{fig:Framework for solving the polynomial of the envelope equation}, the blue box represents the boundary set of the space ${\tilde {\cal D}_i}$, labeled as
\begin{equation} \label{eq:search_space}
    \tilde {\cal D}_i^b = \left\{ {\left. {(\delta \alpha ,\delta \beta )} \right|\delta \alpha  =  \pm 1,\delta \beta  =  \pm 1} \right\} \,.
\end{equation}
The red line is the characteristic curve that consists of the solutions of the envelope equation. The proposed method takes the points on the boundary set $\tilde {\cal D}_i^b$ as initial guesses (\emph{i.e.}, the blue points in Fig.~\ref{fig:Framework for solving the polynomial of the envelope equation}) to search for the solutions (\emph{i.e.}, the red points in Fig.~\ref{fig:Framework for solving the polynomial of the envelope equation}). For the sake of simplicity, the initial guesses are uniformly selected on each bound of the box $\tilde {\cal D}_i^b$ (a box has four bounds: the top, bottom, left, and right bounds). For each initial guess point $(\delta \alpha,\delta \beta ) \in \tilde {\cal D}_i^b$, the coordinate is first represented using the azimuth $\theta$ and distance $r$ as
\begin{equation} \label{eq:theta}
    \theta  = {\tan ^{-1}}\frac{{\delta \alpha }}{{\delta \beta }} \,,
\end{equation}
\begin{equation} \label{eq:r}
    r = \sqrt {\delta {\alpha ^2} + \delta {\beta ^2}} \,.
\end{equation}

Then, we fix the azimuth $\theta$ and solve for the distance $r^{\ast}$ on the ray (\emph{i.e.}, the black dashed line in Fig.~\ref{fig:Framework for solving the polynomial of the envelope equation}) that can satisfy the constraint in Eq.~\eqref{eq:g_constraint}. In this way, the number of variables to be solved, as well as the complexity, are reduced. Let $\boldsymbol{d} = {[\cos \theta,\sin \theta ]^T} \in {\mathbb{R}^2}$, and one has the solution $(\delta {\alpha^{\ast}},\delta {\beta^{\ast}}) = {r^{\ast}}\boldsymbol{d}$, which, after being substituted back into Eqs.~\eqref{eq:g_polynomial}-\eqref{eq:g_constraint}, yields
\begin{equation} \label{eq:g_theta}
    g(\delta {\alpha^{\ast}},\delta {\beta^{\ast}}) \approx {\cal T}_g^{\ast}({r^{\ast}}) = {{\cal T}_g}({r^{\ast}}\sin \theta ,{r^{\ast}}\cos \theta ) = 0 \,.
\end{equation}

Equation~\eqref{eq:g_theta} is a one-variable parameterized polynomial. In this work, the root of Eq.~\eqref{eq:g_theta} is numerically solved using the \emph{fsolve} function in the \emph{Scipy} package \footnote{Available at \url{https://docs.scipy.org/doc/scipy/reference/generated/scipy.optimize.fsolve.html}.}. Note that the root $r^{\ast}$ has both lower and upper bounds, as it should be inside the search space ${\tilde {\cal D}_i}$. The lower bound for $r^{\ast}$ is 0, whereas the upper bound is ${r^{\ast}} \le r$, depending on the location of the initial guess $(\delta \alpha,\delta \beta ) \in \tilde {\cal D}_i^b$. For example, when $\theta = \frac{\pi}{4}$, ${r^{\ast}} \le \sqrt 2 $, and when $\theta = 0$, ${r^{\ast}} \le 1 $. If no feasible root is found for the given search direction, the corresponding initial guess is considered as a point on the characteristic curve that determines the RS boundary of the \emph{i}-th sub-domain ${{\cal D}_i}$. This means that the boundary yielded here is due to the limitation of the search space rather than the two-variable parameterized curve (\emph{i.e.}, Eqs.~\eqref{eq:xp_polynomial}-\eqref{eq:zp_polynomial}) having an extreme point along the given direction. In other words, if the search space ${\tilde {\cal D}_i}$ is larger, the RS boundary will move.

It should be noted that the above method is valid only when there is a unique solution to the envelope equation along a given ray (\emph{i.e.}, the azimuth $\theta$). The use of ADS technology to split the initial domain into multiple subdomains ensures that the RS’s envelope remains relatively simple within each subdomain. Moreover, our simulations have not observed the presence of multiple solutions (see Sec.~\ref{Sec:Numerical simulations}). However, we cannot rigorously prove that multiple solutions never exist along a given ray. In cases where multiple solutions may occur, one potential approach to mitigate this issue is to reduce the ADS threshold, resulting in a finer split of the initial domain. This refinement decreases the likelihood that multiple solutions appear. Another potential way to address multiple solutions is to select the solution that is closest to the guess point (\emph{i.e.}, the blue points in Fig.~\ref{fig:Framework for solving the polynomial of the envelope equation}).

\subsection{Local Polynomial Approximation of the Envelope}
\label{Sec:Local Polynomial Approximation of the Envelope}
Although it is very fast to search for the root (as Eq.~\eqref{eq:g_theta} is analytical), its computational cost grows linearly with the number of initial guesses. Therefore, it is still inefficient if we want to identify the envelope points densely. For example, if we sample 51 points on each bound of the search space ${\tilde {\cal D}_i}$, we need to solve the equation 200 times after removing the repeated points (the corner of the square).

\begin{figure}[!h]
    \centering
    \includegraphics[width=0.7\linewidth]{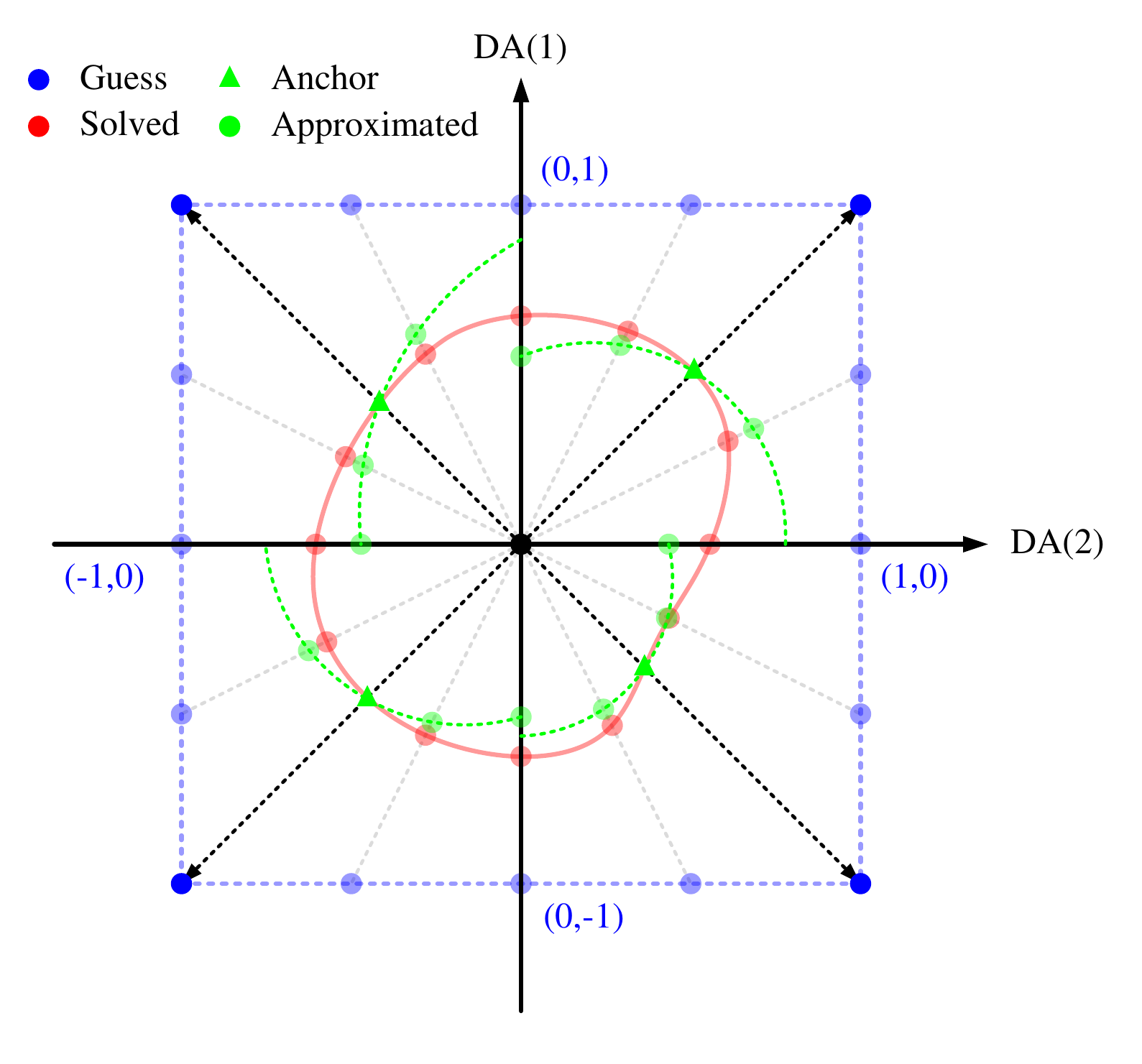}
    \caption{\label{fig:Analytical pieces for solving the envelope equation} Analytical pieces for solving the envelope equation.}
\end{figure}

To improve computational efficiency, a method has been developed to approximate the roots analytically instead of searching them numerically. As shown in Fig.~\ref{fig:Analytical pieces for solving the envelope equation}, the envelope equation is numerically solved only on a few points (noted as the anchor points), and the roots of the envelope equation on other points are approximated (green points) using analytical pieces (noted as local polynomial approximations) expanded from these anchor points.

Let $({\theta _k},r_k^{\ast})$ be the pair of azimuth and the solved distance of the \emph{k}-th anchor point. By adding variations $\delta {\theta _k}$ and $\delta r_k^{\ast}$ as DA variables and using Eqs.~\eqref{eq:theta}-\eqref{eq:r}, one has
\begin{equation} \label{eq:dalpha_polynomial}
    \delta \alpha  \approx {{\cal T}_{\delta \alpha }}(\delta {\theta _k},\delta r_k^{\ast}) \,,
\end{equation}
\begin{equation} \label{eq:dbeta_polynomial}
    \delta \beta  \approx {{\cal T}_{\delta \beta }}(\delta {\theta _k},\delta r_k^{\ast}) \,.
\end{equation}
Substitution of Eqs.~\eqref{eq:dalpha_polynomial}-\eqref{eq:dbeta_polynomial} back into Eq.~\eqref{eq:g_constraint} yields
\begin{equation} \label{eq:g_dtheta}
    g \approx {{\cal T}_g}(\delta {\theta _k},\delta r_k^{\ast}) = 0 \,,
\end{equation}
which, after being augmented by adding an identified map $\delta {\theta _k} = {{\cal I}_{\delta {\theta _k}}}(\delta {\theta _k})$, becomes
\begin{equation} \label{eq:g_dtheta_augmented}
    \left[ {\begin{array}{*{20}{c}}
        {\delta {\theta _k}}\\
        g
        \end{array}} \right] = \left[ {\begin{array}{*{20}{c}}
            {{{\cal I}_{\delta {\theta _k}}}}\\
            {{{\cal T}_g}}
        \end{array}} \right] \cdot \left[ {\begin{array}{*{20}{c}}
        {\delta {\theta _k}}\\
        {\delta r_k^{\ast}}
    \end{array}} \right] \,.
\end{equation}

Inverting the augmented map in Eq.~\eqref{eq:g_dtheta_augmented}, enforcing $g = 0$, one can derive the local polynomial approximation as
\begin{equation} \label{eq:local_polynomial}
    \delta r_k^{\ast} \approx {{\cal T}_{\delta r_k^{\ast}}}(\delta {\theta _k}) \,,
\end{equation}
which is an analytical polynomial piece that expresses the effects of a variation of $\theta_{k}$ on the solution of the envelope equation as a high-order polynomial. Thus, for a point neighboring the given anchor point, the solution is approximated using the local polynomial approximation as
\begin{equation} \label{eq:local_polynomial_solution}
    \begin{aligned}
        & \theta \approx {\theta _k} + \delta {\theta _k} \\
        & r^{\ast} \approx r_k^{\ast} + {{\cal T}_{\delta r_k^{\ast}}}(\delta {\theta _k})
    \end{aligned} \,.
\end{equation}

\begin{figure}[!h]
    \centering
    \includegraphics[width=0.85\linewidth]{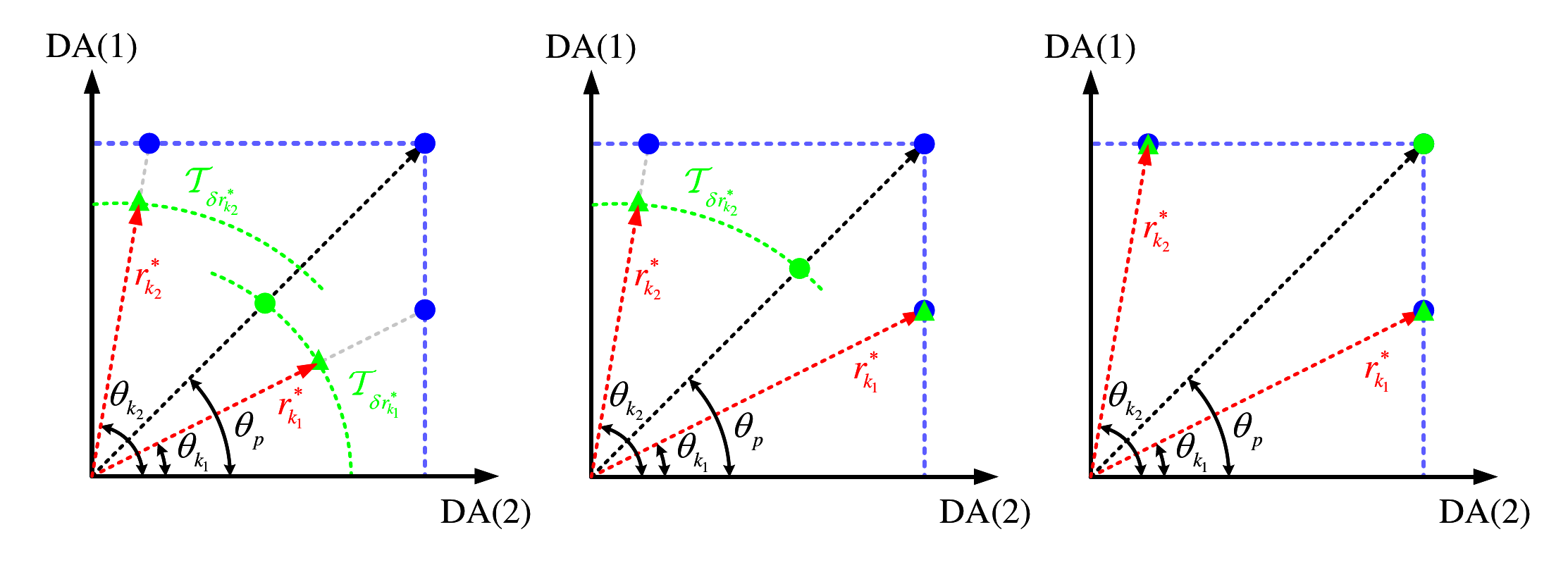}
    \caption{\label{fig:Strategy of choosing analytical pieces} Strategy of choosing analytical pieces.}
\end{figure}

As the anchor point may not have a feasible root, such that it is located on the bounds of the search space ${\tilde {\cal D}_i}$, solutions are provided to address these cases herein. Assume that $N_a$ anchor points ($({\theta _k},r_k^{\ast}),k \in \{ 1, \cdots ,{N_a}\} $) are solved. As shown in Fig.~\ref{fig:Strategy of choosing analytical pieces}, given a new azimuth $\theta_p$, one can find the nearest anchor points (from the angle of azimuth) on both sides, labeled as $({\theta _{{k_1}}},r_{{k_1}}^{\ast})$ and $({\theta _{{k_2}}},r_{{k_2}}^{\ast})$. If both anchor points have feasible roots, choose the local polynomial approximation of the closest one to predict the root $r_p^{\ast}$ (see the left sub-figure of Fig.~\ref{fig:Strategy of choosing analytical pieces}). If only one anchor point has a feasible root, one should choose the sole local polynomial approximation (see the middle sub-figure of Fig.~\ref{fig:Strategy of choosing analytical pieces}). If both anchor points do not have a feasible root, the envelope equation is considered to have no feasible root for the given azimuth (see the right sub-figure of Fig.~\ref{fig:Strategy of choosing analytical pieces}), and the corresponding initial guess becomes a characteristic point. 

Note that it’s also possible that the solution approximated by the local polynomial approximation (\emph{i.e.}, Eq.~\eqref{eq:local_polynomial_solution}) exceeds its lower or upper bounds (\emph{i.e.}, $r_k^{\ast} + {{\cal T}_{\delta r_k^{\ast}}}(\delta {\theta _k}) < 0$ or $r_k^{\ast} + {{\cal T}_{\delta r_k^{\ast}}}(\delta {\theta _k}) > r_k$). In this case, that root is rejected, and its initial guess (blue points in Fig.~\ref{fig:Strategy of choosing analytical pieces}) is chosen as a characteristic point.

The above steps determine the characteristic curve of the \emph{i}-th sub-domain ${{\cal D}_i}$. By evaluating the points on the characteristic curve using the polynomials in Eqs.~\eqref{eq:xp_polynomial}-\eqref{eq:zp_polynomial}, the envelope of the RS of \emph{i}-th sub-domain ${{\cal D}_i}$ can be obtained. In this paper, the envelopes of the sub-domains are called “sub-envelopes”. Note that the sub-envelopes of different sub-domains may overlap. In other words, the sub-envelope points of one sub-domain may be the interior points of another sub-domain. Thus, we need to identify the final envelope and delete “interior” sub-envelope points after obtaining all the sub-envelope points. In this work, the alpha-shape method \footnote{Using Python package \emph{alphashape}: available at \url{https://pypi.org/project/alphashape/}.}, which is a geometric technique used to analyze and describe the shape of a set of points in a multi-dimensional space, is employed to identify final envelopes and delete interior points. It extends the concept of the convex hull by allowing for the extraction of more complex shapes that can better represent the topology of the points. 

Lastly, note that the procedures described above can also be applied to determine the boundary for the RS in observation space by simply replacing terms in the envelope equation (\emph{i.e.}, Eqs.~\eqref{eq:xp_polynomial}-\eqref{eq:zp_polynomial}) with Eqs.~\eqref{eq:gamma_polynomial}-\eqref{eq:delta_polynomial}.

\subsection{Overall Procedure} 
\label{Sec:Overall Procedure}
Based on the discussions in Sec.~\ref{Sec:Reachable Set in State Space}, Sec.~\ref{Sec:Framework for Solving the Envelope Equation} and Sec.~\ref{Sec:Local Polynomial Approximation of the Envelope}, the overall procedure of the algorithm to determine the final envelope of the RS in the state space is shown in Table~\ref{tab:Pseudocode to determine the boundary of RS in state space}. Recall that the RS in observation space can be determined by replacing the steps 1-3 in Table~\ref{tab:Pseudocode to determine the boundary of RS in state space} with Eqs.~\eqref{eq:gamma_polynomial}-\eqref{eq:delta_polynomial}.

\begin{table}[!h]
    \caption{\label{tab:Pseudocode to determine the boundary of RS in state space} Pseudocode to determine the boundary of RS in state space}
    \centering
    \begin{tabular}{lr}
        \hline\hline
        \multicolumn{2}{l}{\makecell[l]{\textbf{Input}: maximal velocity impulse $\Delta v_{\max}$, orbital dynamics $\boldsymbol{f}(\boldsymbol{x})$, initial epoch $t_0$, final epoch $t_f$, threshold $\epsilon$, \\ maximum order of the polynomials $n$, initial domain ${\cal D}$, number of initial guess points per bound $N_p$, number \\ of anchor points per bound $N_a$.}} \\
        \multicolumn{2}{l}{\textbf{Output}: the final envelope of the RS.} \\ \hline
        \,\,\,\,\textbf{1}: Determine the transformation matrix $\boldsymbol{T}(\boldsymbol{\bar x})$ at the final epoch $t_f$. & $\rhd$ Eq.~\eqref{eq:Tx} \\
        \,\,\,\,\textbf{2}: Derive the polynomial ${\boldsymbol{r}_p} \approx {{\cal T}_{{\boldsymbol{r}_p}}}(\alpha ,\beta ,\delta t)$ at the final epoch $t_f$. & $\rhd$ Eqs.~\eqref{eq:rp}-\eqref{eq:rp_t_polynomial} \\
        \makecell[l]{\,\,\,\,\textbf{3}: Perform partial map inversion (or Newton's iteration) to obtain ${x_p} \approx {{\cal T}_{{x_p}}}(\alpha ,\beta )$ \\\,\,\,\,\,\,\,\,\,\,\,and ${z_p} \approx {{\cal T}_{{z_p}}}(\alpha ,\beta )$.} & $\rhd$ Eqs.~\eqref{eq:dyp_polynomial}-\eqref{eq:tf_correction} \\
        \,\,\,\,\textbf{4}: Split the initial domain ${\cal D}$ using the ADS to obtain sub-domains ${\cal D}_i$ ($i \in \{ 1, \cdots ,N\} $). &  \\
        \,\,\,\,\textbf{5}: Derive the polynomial for the envelope equation for each sub-domain. & $\rhd$ Eqs.~\eqref{eq:g}-\eqref{eq:g_polynomial} \\
        \,\,\,\,\textbf{6}: Generate initial guess points and anchor points for each sub-domain. & \\
        \,\,\,\,\textbf{7}: Solve the envelope equations for the anchor points. & $\rhd$ Eqs.~\eqref{eq:g_constraint}-\eqref{eq:g_theta} \\
        \,\,\,\,\textbf{8}: Derive the local polynomial approximation based on each anchor point. & $\rhd$ Eqs.~\eqref{eq:dalpha_polynomial}-\eqref{eq:local_polynomial} \\
        \,\,\,\,\textbf{9}: Predict the solution using the local polynomial approximation for each initial guess point. & $\rhd$ Eq.~\eqref{eq:local_polynomial_solution} \\
        \,\,\,\,\textbf{10}: Generate characteristic curves and determine sub-envelopes for each sub-domain. & $\rhd$ Eqs.~\eqref{eq:xp_polynomial}-\eqref{eq:zp_polynomial} \\
        \,\,\,\,\textbf{11}: Determine the final envelope by combing all the sub-envelopes. & \\
        \hline\hline
    \end{tabular}
\end{table}

Among the inputs of the algorithm, the maximal velocity impulse $\Delta v_{\max}$, orbital dynamics $\boldsymbol{f}(\boldsymbol{x})$, initial epoch $t_0$ and final epoch $t_f$ are determined by the simulation task, whereas the threshold $\epsilon$, maximum order $n$, and two numbers $N_p$ and $N_a$ are user-defined parameters. Note that the result of the proposed algorithm is an approximated envelope of the RS rather than an exact one. This is because our algorithm approximates the orbital motion using high-order polynomials. The precision of the identified envelope relies on two user-defined parameters: the threshold $\epsilon$ and the maximum order $n$. Both decreasing the threshold $\epsilon$ and increasing the maximal order $n$ can improve accuracy. In addition, increasing the number of anchor points can improve the accuracy of the local polynomial approximations, which in turn reduces the errors of the final envelope. However, all the above three efforts require a higher computational cost.

\section{Numerical simulations}
\label{Sec:Numerical simulations}
The proposed method is applied to determine the RS of orbits in cislunar space, where the dynamics are highly nonlinear and lack an analytical orbital propagation solution. In this case, the previous RS methods are inapplicable. Simulations are implemented on a personal laptop with a 2.5 GHz processor (12th Gen Intel(R) Core(TM) i5-12500H) and 16 GB RAM.

\subsection{Scenario Design} 
\label{Sec:Scenario Design}
The CRTBP dynamics is employed in the numerical simulation, with the mathematical model expressed as
\begin{equation} \label{eq:CRTBP}
    \left\{ \begin{array}{l}
        \ddot x = 2\dot y + x - \frac{{(1 - \mu )(x + \mu )}}{{r_1^3}} - \frac{{\mu (x + \mu  - 1)}}{{r_2^3}}\\
        \ddot y =  - 2\dot x + y - \frac{{(1 - \mu )y}}{{r_1^3}} - \frac{{\mu y}}{{r_2^3}}\\
        \ddot z =  - \frac{{(1 - \mu )z}}{{r_1^3}} - \frac{{\mu z}}{{r_2^3}}
    \end{array} \right. \,,
\end{equation}
where $\boldsymbol{x} = [\boldsymbol{r};\boldsymbol{v}] = [x,y,z,\dot x,\dot y,\dot z]^T \in {\mathbb{R}^6}$ is the non-dimensional orbital state in the Earth-Moon rotating frame, $\mu=0.0121505839$ is the non-dimensional gravitational constant of the Earth-Monn three-body system, and 
\begin{equation} \label{eq:r1}
    {r_1} = \sqrt {{{(x + \mu )}^2} + {y^2} + {z^2}} \,,
\end{equation}
\begin{equation} \label{eq:r2}
    {r_2} = \sqrt {{{(x + \mu  - 1)}^2} + {y^2} + {z^2}} \,.
\end{equation}

Two NRHOs, a linearly stable NRHO and a 9:2 NRHO (for the Gateway mission), are simulated, with their initial orbital states listed in Table~\ref{tab:Initial orbital state of the linearly stable NRHO} and Table~\ref{tab:Initial orbital state of the 9:2 NRHO}. These two NRHOs are in the vicinity of the Earth-Moon L2 Lagrange point. As non-dimensional coordinates are employed, the length unit (LU), velocity unit (VU), and time unit (TU) are provided in Table 2. The linearly stable NRHO starts at its apolune (\emph{i.e.}, initial epoch) and has a period (labeled as $T$) of 2.667 TU (approximately 9.8435 days). The 9:2 NRHO also starts at its apolune, with its orbital period being 1.5112 TU (approximately 6.5624 days). Compared to the linearly stable NRHO, the 9:2 NRHO is much closer to the Moon at its perilune \cite{Fu2024}. Thus, its motion is highly nonlinear at that point, leading to a numerical singularity when implementing the map inversion.

\begin{table}[!h]
    \caption{\label{tab:Initial orbital state of the linearly stable NRHO} Initial orbital state of the linearly stable NRHO}
    \centering
    \begin{tabular}{lrl}
        \hline\hline
        \multicolumn{2}{l}{Parameter}                     & Value                \\ \hline
        \multirow{3}{*}{Position vector (nd)} & $x$       & 1.07523949148639     \\
                                              & $y$       & 0                    \\
                                              & $z$       & -0.202146176080457   \\ \hline
        \multirow{3}{*}{Velocity vector (nd)} & $\dot{x}$ & 0                    \\
                                              & $\dot{y}$ & -0.192431661980241   \\
                                              & $\dot{z}$ & 0                    \\ \hline
        \multicolumn{2}{l}{Length unit (km)}              & 384400               \\ \hline
        \multicolumn{2}{l}{Velocity unit (km/s)}          & 1.02454629434750     \\ \hline
        \multicolumn{2}{l}{Time unit (s)}                 & 375190.464423878     \\ \hline
        \multicolumn{2}{l}{Period (nd)}                   & 2.26679784217712     \\
        \hline\hline
    \end{tabular}
\end{table}

\begin{table}[!h]
    \caption{\label{tab:Initial orbital state of the 9:2 NRHO} Initial orbital state of the 9:2 NRHO}
    \centering
    \begin{tabular}{lrl}
        \hline\hline
        \multicolumn{2}{l}{Parameter}                     & Value                \\ \hline
        \multirow{3}{*}{Position vector (nd)} & $x$       & 1.02202815472411     \\
                                              & $y$       & 0                    \\
                                              & $z$       & -0.182101352652963   \\ \hline
        \multirow{3}{*}{Velocity vector (nd)} & $\dot{x}$ & 0                    \\
                                              & $\dot{y}$ & -0.103270818092086   \\
                                              & $\dot{z}$ & 0                    \\ \hline
        \multicolumn{2}{l}{Period (nd)}                   & 1.51119865689808     \\
        \hline\hline
    \end{tabular}
\end{table}

In the numerical simulations, the values of those user-defined parameters are listed in Table~\ref{tab:User-defined parameters for simulations}. A total of 20 anchor points are to be solved for each sub-domain, whereas the remaining 200 points are approximated using the local polynomial approximation. The threshold $\epsilon$ that controls the errors of the polynomials is set differently in the two examples (\emph{i.e.}, the state and observation RS examples), which will be discussed in the following two subsections.

\begin{table}[!h]
    \caption{\label{tab:User-defined parameters for simulations} User-defined parameters for simulations}
    \centering
    \begin{tabular}{lll}
        \hline\hline
        Parameter & Symbol & Value \\ \hline
        Maximal velocity impulse & $\Delta v_{\max}$ & 10 m/s \\
        Maximal order of the polynomials & $n$ & 6 \\
        Number of initial guess points per bound & $N_p$ & 51 \\
        Number of anchor points per bound & $N_a$ & 6 \\
        Threshold & $\epsilon$ & - \\
        \hline\hline
    \end{tabular}
\end{table}

Three numerical examples are presented in the following subsections. The first two examples are RS in state space, with the first based on the linearly stable NRHO (Sec.~\ref{Sec:Reachable Set in State Space: Linearly Stable NRHO Case}) and the second based on the 9:2 NRHO (Sec.~\ref{Sec:Reachable Set in State Space: 9:2 NRHO Case}). For the first example, the RS is modeled using the partial map inversion approach, whereas, for 9:2 NRHO, the inversion approach is infeasible due to the nonlinearity near the perilune. Additionally, the third example shows the RS in observation (Sec.~\ref{Sec:Reachable Set in Observation Space (simulation)}), where an observer on the 9:2 NRHO is assumed to observe the target on the linearly stable NRHO.

\subsection{Reachable Set in State Space: Linearly Stable NRHO Case}
\label{Sec:Reachable Set in State Space: Linearly Stable NRHO Case}
\begin{figure}[!h]
    \centering
    \includegraphics[width=0.48\linewidth]{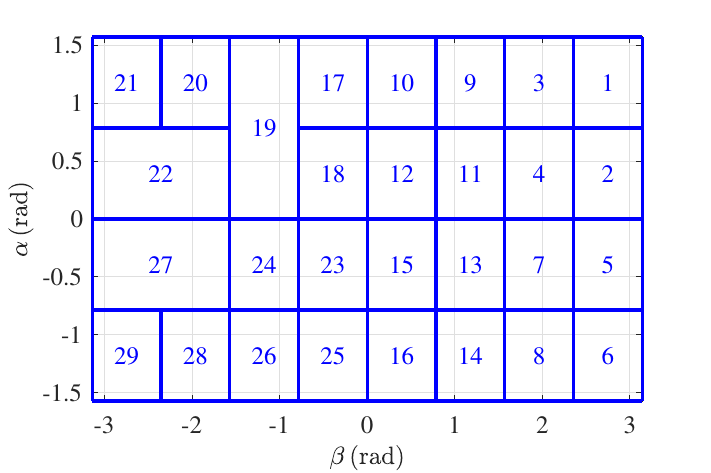}
    \caption{\label{fig:Split Sub-domains of the state case} Split Sub-domains of the state case.}
\end{figure}

\begin{figure}[!h]
    \centering
    \subfigure[Initial guesses, anchor points, and approximated characteristic points]{
        \label{fig:8a}
        \includegraphics[width=0.48\linewidth]{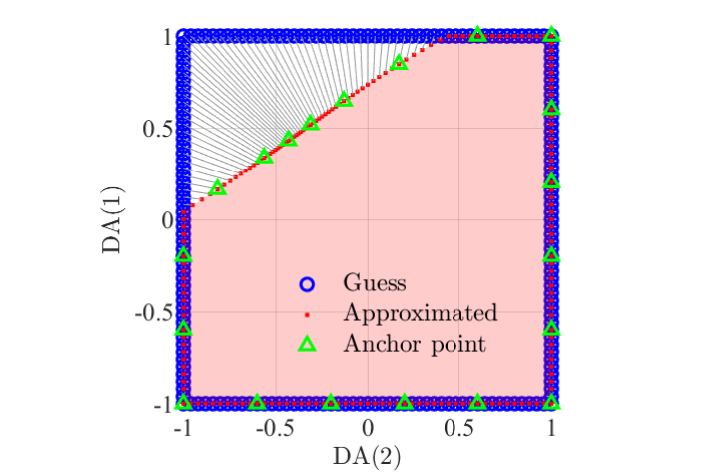}
    }
    \subfigure[Approximated and solved (exact) characteristic points]{
        \label{fig:8b}
        \includegraphics[width=0.48\linewidth]{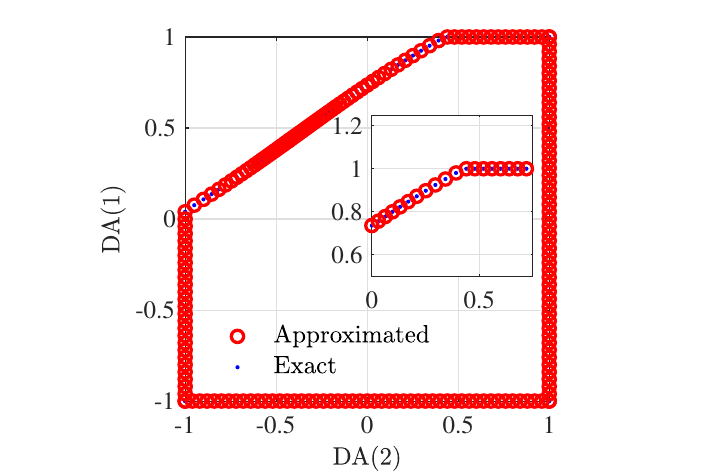}
    }
    \caption{\label{fig:Characteristic curve of the 24th sub-domain} Characteristic curve of the $24^{\rm{th}}$ sub-domain.}
\end{figure}

For this case, the first approach, say, the partial map inversion, is employed to derive the polynomials for the states (\emph{i.e.}, $x_p$ and $z_p$) on the auxiliary plane.
First, starting from the apolune point, the RS after one orbital period’s propagation is determined using the proposed method (\emph{i.e.}, ${t_f} = 1T \approx 2.667{\rm{ (nd)}}$). The threshold of this simulation is set as $\epsilon=10^{-5}$. Using the ADS, the initial domain ${\cal D}$ is divided into 29 sub-domains ${{\cal D}_i}$ ($i \in \{ 1,2, \cdots ,29\} $). Figure~\ref{fig:Split Sub-domains of the state case} demonstrates all sub-domains distinguished by their IDs. Figure~\ref{fig:Characteristic curve of the 24th sub-domain} shows the characteristic curve of the $24^{\rm{th}}$ sub-domain (\emph{i.e.}, ${\tilde {\cal D}_{24}}$). In Fig.~\ref{fig:Characteristic curve of the 24th sub-domain}\subref{fig:8a}, the initial guesses (boundary points in the set $\tilde {\cal D}_{24}^b$), the anchor points, and the characteristic points approximated by the local polynomial approximations are represented using blue circles, green triangles, and red dots, respectively. The initial guesses and the corresponding characteristic points are paired using gray lines. Additionally, Fig.~\ref{fig:Characteristic curve of the 24th sub-domain}\subref{fig:8b} compares the characteristic points approximated by the local polynomial approximations (labeled as “approximated”) and the ones obtained by numerically solving the envelope equation (labeled as “solved”). One can see from Fig.~\ref{fig:Characteristic curve of the 24th sub-domain}\subref{fig:8b} that the approximated points (red circles) are in good agreement with the solved data (blue dots), with absolute errors at the level of $10^{-12}$. To show the benefits of using local polynomial approximations in terms of computational efficiency, the CPU time is recorded and compared. If all the characteristic points (for all 29 sub-domains) are obtained by numerically solving the envelope equation, the CPU time is 12.3184 s, whereas using the anchor points and local polynomial approximations, the CPU time is 1.9497 s (including solving the envelope equation for the anchor points, and constructing and using the local polynomial approximations). A reduction of 84.17\% in computational time can be benefited from the local polynomial approximations. That percentage of reduction will still increase if more characteristic points are approximated or fewer anchor points are used.

\begin{figure}[!h]
    \centering
    \subfigure[Sub-envelope]{
        \label{fig:9a}
        \includegraphics[width=0.48\linewidth]{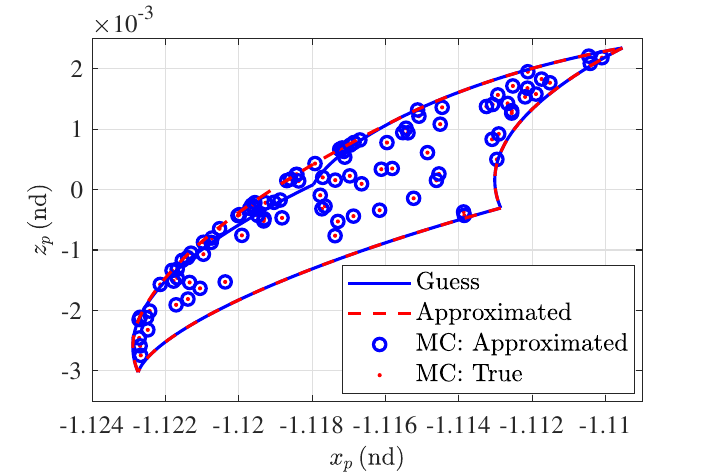}
    }
    \subfigure[Larger plot of \subref{fig:9a}]{
        \label{fig:9b}
        \includegraphics[width=0.48\linewidth]{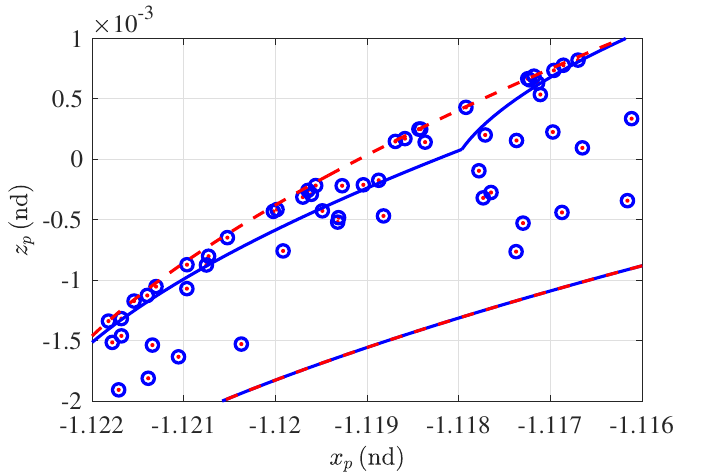}
    }
    \caption{\label{fig:Sub-envelope of the 24th sub-domain} Sub-envelope of the $24^{\rm{th}}$ sub-domain.}
\end{figure}

Figure~\ref{fig:Sub-envelope of the 24th sub-domain}\subref{fig:9a} shows the sub-envelope of the $24^{\rm{th}}$ sub-domain on the $x_p$-$z_p$ plane, and Fig.~\ref{fig:Sub-envelope of the 24th sub-domain}\subref{fig:9b} demonstrates where the gaussed and approximated sub-envelopes mismatches. In Fig.~\ref{fig:Sub-envelope of the 24th sub-domain}, the blue and red lines are the initial guesses and predictions of the sub-envelopes, which are obtained by directly evaluating the polynomials in Eqs.~\eqref{eq:xp_polynomial}-\eqref{eq:zp_polynomial} using the points on the boundary of the sub-domain (\emph{i.e.}, the blue circles in Fig.~\ref{fig:Characteristic curve of the 24th sub-domain}\subref{fig:8a}) and the characteristic points (\emph{i.e.}, the red dots in Fig.~\ref{fig:Characteristic curve of the 24th sub-domain}\subref{fig:8a}), respectively. Additionally, 100 MC points are presented to validate the accuracy of the obtained sub-envelope. For each MC run, an impulsive maneuver is randomly generated inside the $24^{\rm{th}}$ sub-domain. By adding the velocity impulse to the nominal state, propagating the trajectory, and transforming the coordinates, the $x_p$-$z_p$ pairs of 100 MC runs are shown by red dots in Fig.~\ref{fig:Characteristic curve of the 24th sub-domain}. As a comparison, the $x_p$-$z_p$ pairs are also approximated using the high-order polynomials and are shown by blue circles. It can be seen from Fig.~\ref{fig:Characteristic curve of the 24th sub-domain} that the red dots and the blue circles overlap, indicating that the polynomials can accurately predict the $x_p$-$z_p$ pairs in the given sub-domain. Moreover, as shown in Fig.~\ref{fig:Sub-envelope of the 24th sub-domain}\subref{fig:9b}, some of the MC points are located outside the guessed boundary (blue lines); however, they are all covered by the approximated sub-envelope (the red lines). 

\begin{figure}[!h]
    \centering
    \subfigure[Characteristic curve]{
        \label{fig:10a}
        \includegraphics[width=0.48\linewidth]{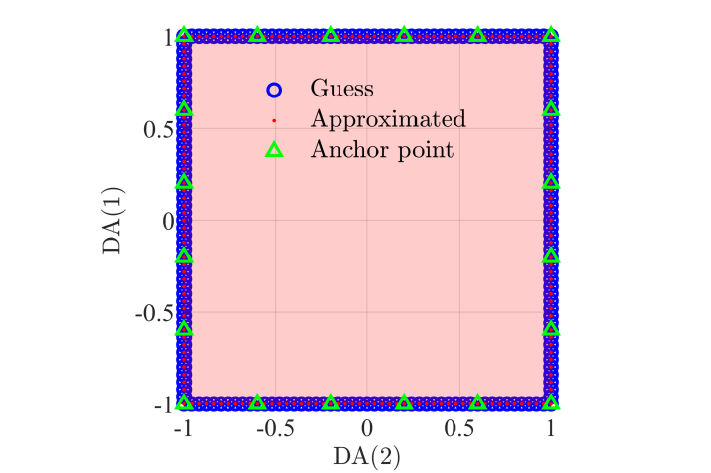}
    }
    \subfigure[Sub-envelope]{
        \label{fig:10b}
        \includegraphics[width=0.48\linewidth]{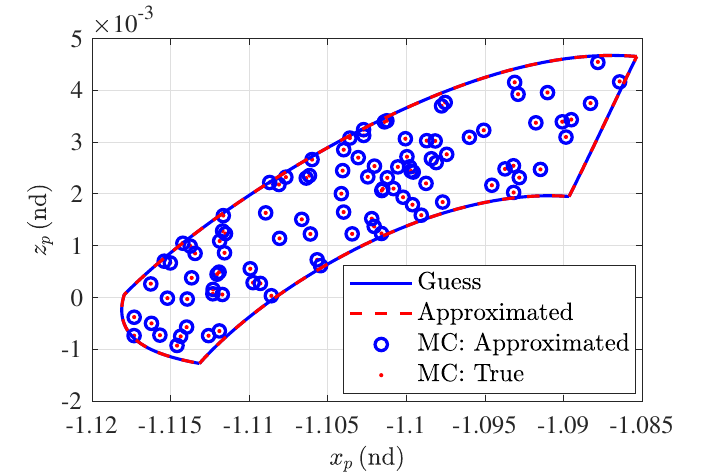}
    }
    \caption{\label{fig:Results of the 22nd sub-domain} Results of the $22^{\rm{nd}}$ sub-domain.}
\end{figure}

\begin{figure}[!h]
    \centering
    \subfigure[Characteristic curve]{
        \label{fig:11a}
        \includegraphics[width=0.48\linewidth]{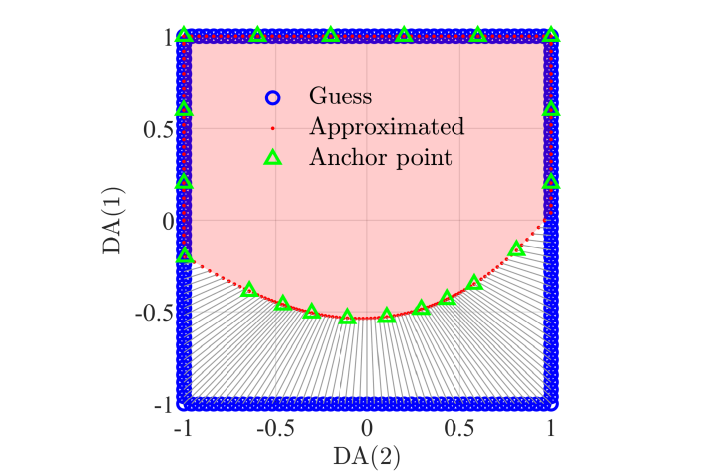}
    }
    \subfigure[Sub-envelope]{
        \label{fig:11b}
        \includegraphics[width=0.48\linewidth]{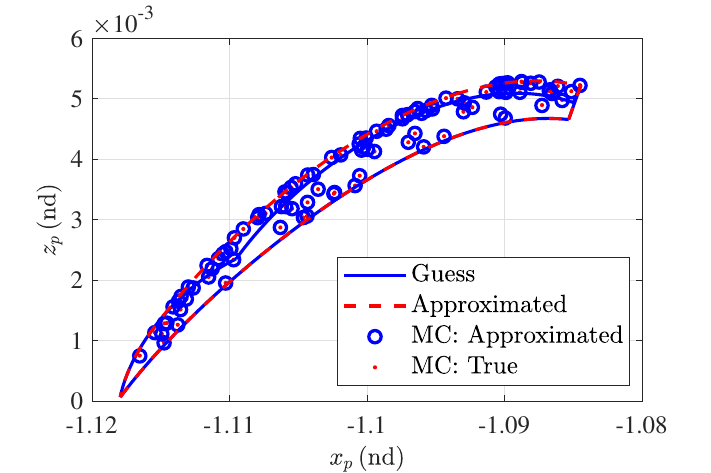}
    }
    \caption{\label{fig:Results of the 27th sub-domain} Results of the $27^{\rm{th}}$ sub-domain.}
\end{figure}

Figure~\ref{fig:Results of the 22nd sub-domain} and Fig.~\ref{fig:Results of the 27th sub-domain} provide the sub-envelopes of the $22^{\rm{nd}}$ and $27^{\rm{th}}$ sub-domains, respectively. The $22^{\rm{nd}}$ sub-domain is a special case as all its characteristic points are located on its boundary. As shown in Fig.~\ref{fig:Results of the 22nd sub-domain}\subref{fig:10b} and Fig.~\ref{fig:Results of the 27th sub-domain}\subref{fig:11b}, all the MC points are accurately approximated and are wrapped by the approximated sub-envelopes.

\begin{figure}[!h]
    \centering
    \includegraphics[width=0.48\linewidth]{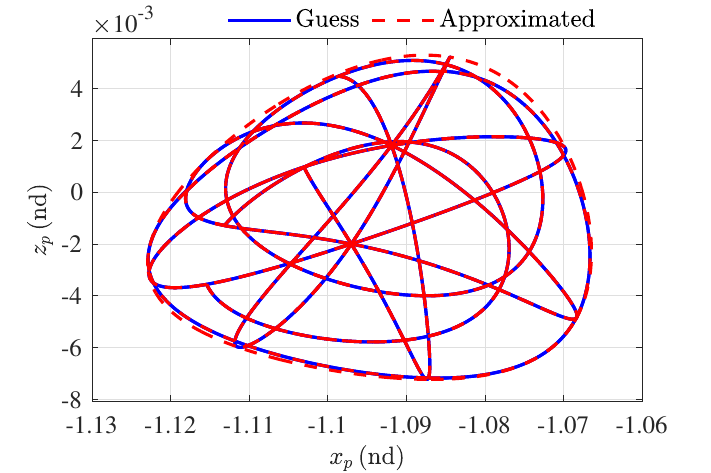}
    \caption{\label{fig:Sub-envelopes of all sub-domains of the state case} Sub-envelopes of all sub-domains of the state case.}
\end{figure}

\begin{figure}[!h]
    \centering
    \includegraphics[width=0.48\linewidth]{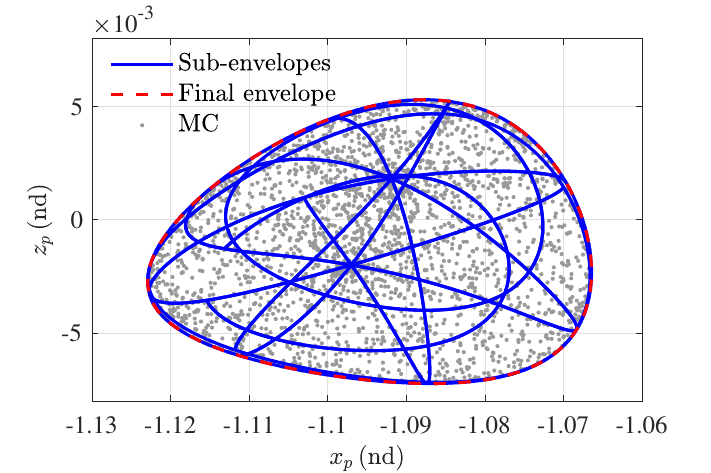}
    \caption{\label{fig:Final envelope of the state case} Final envelope of the state case.}
\end{figure}

\begin{figure}[!h]
    \centering
    \subfigure[3D plots]{
        \label{fig:14a}
        \includegraphics[width=0.75\linewidth]{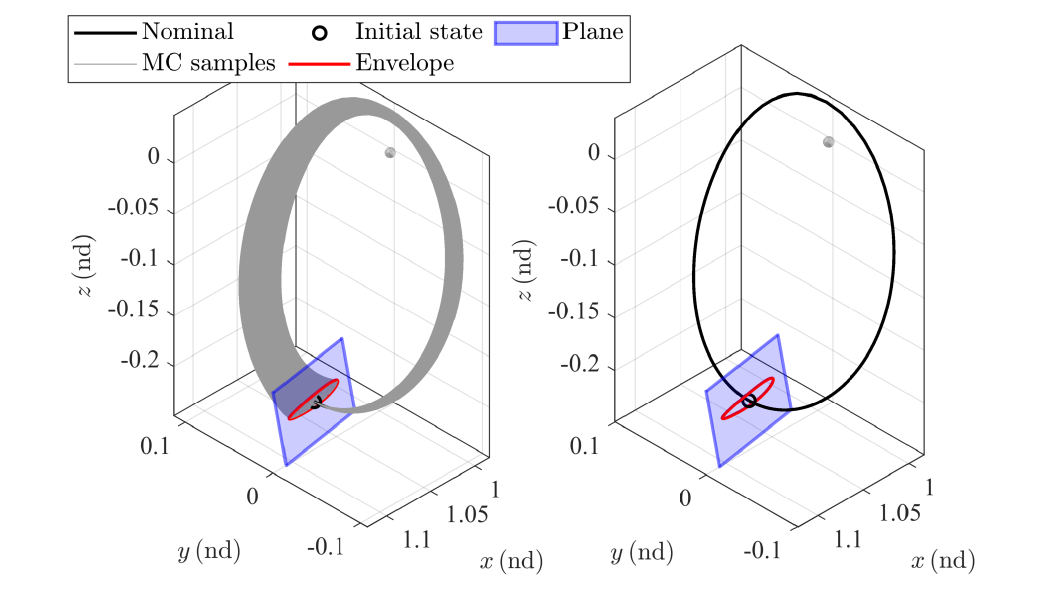}
    }
    \subfigure[Larger plot of \subref{fig:14a}]{
        \label{fig:14b}
        \includegraphics[width=0.75\linewidth]{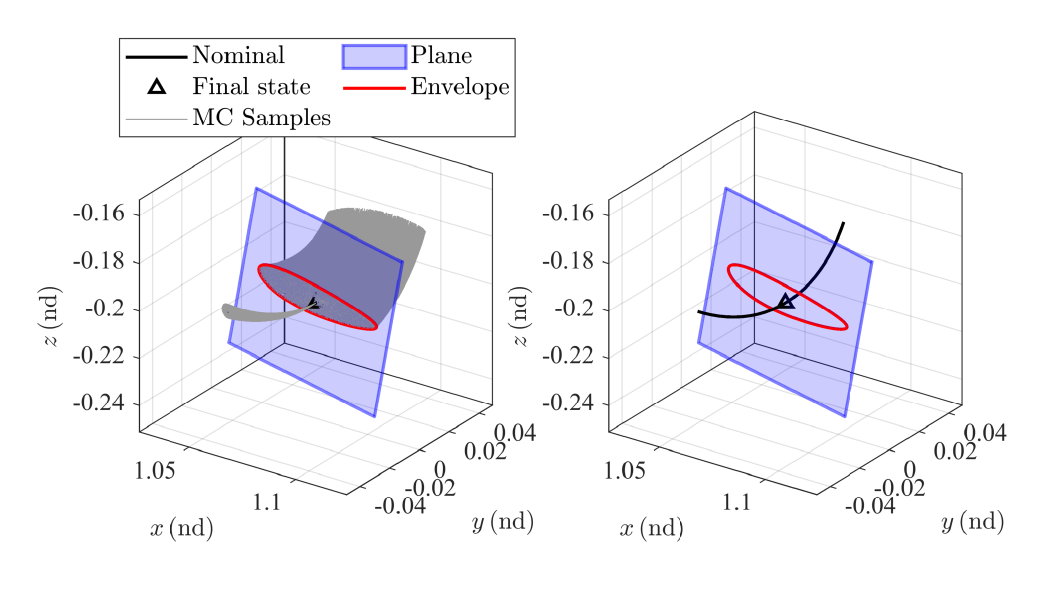}
    }
    \caption{\label{fig:RS projected on the characterized plane} RS projected on the auxiliary plane.}
\end{figure}

Figure~\ref{fig:Sub-envelopes of all sub-domains of the state case} compares the guessed and approximated sub-envelopes of all 29 sub-domains. By combining these sub-envelopes using the alpha-shape method, the final envelope is shown by the red dashed line in Fig.~\ref{fig:Final envelope of the state case}. Figure~\ref{fig:RS projected on the characterized plane} gives a 3D plot of the RS, with the auxiliary plane and the final envelope colored blue and red, respectively. Additionally, 2900 MC points (100 points for each sub-domain) are added to Fig.~\ref{fig:Final envelope of the state case} and Fig.~\ref{fig:RS projected on the characterized plane} to show the accuracy of the obtained final envelope. The accuracy of the RS envelope is verified by comparing it with the MC points. Motivated by Ref.~\cite{Wen2022JGCD}, an error index
\begin{equation} \label{eq:error_index}
    P = \frac{{d_{\max}^2}}{{{S_{{\rm{RS}}}}}} \times 100\% 
\end{equation}
is employed to quantitatively evaluate the relative error of the RS envelope, where $S_{\rm{RS}}$ is the area of the RS and $d_{\max}$ is the maximal distance between the RS envelope and the MC points that are outside the RS envelope. For this case, $S_{\rm{RS}}=5.4043 \times 10^{-4}\,(\rm{nd}^2)$ and $d_{\max}=6.3388 \times 10^{-6}\,(\rm{nd})$; thus, the relative error is $P=7.4350 \times 10^{-8}$.

\begin{figure}[!h]
    \centering
    \includegraphics[width=0.48\linewidth]{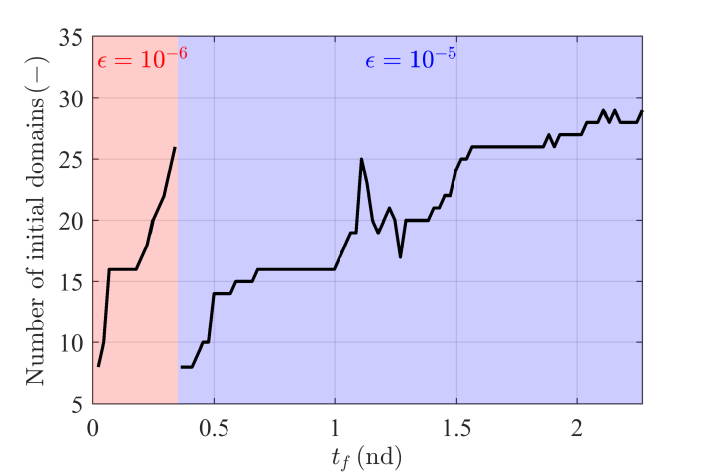}
    \caption{\label{fig:Time history of the number of sub-domains of the state case} Time history of the number of sub-domains of the state case.}
\end{figure}

\begin{figure}[!h]
    \centering
    \includegraphics[width=0.75\linewidth]{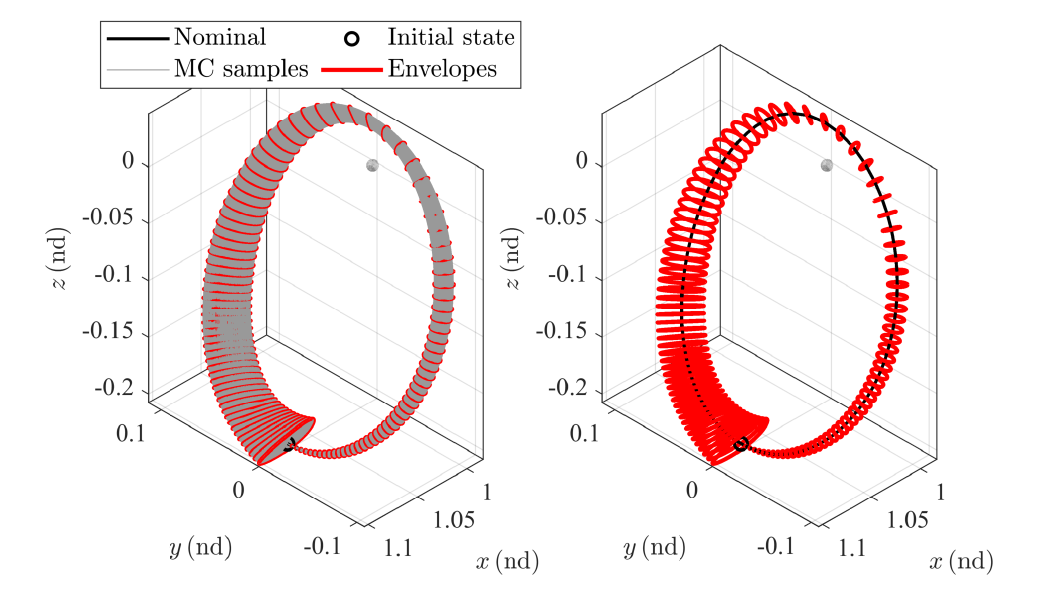}
    \caption{\label{fig:Envelopes of the state RSs at different epochs} Envelopes of the state RSs at different epochs.}
\end{figure}

\begin{figure}[!h]
    \centering
    \includegraphics[width=0.9\linewidth]{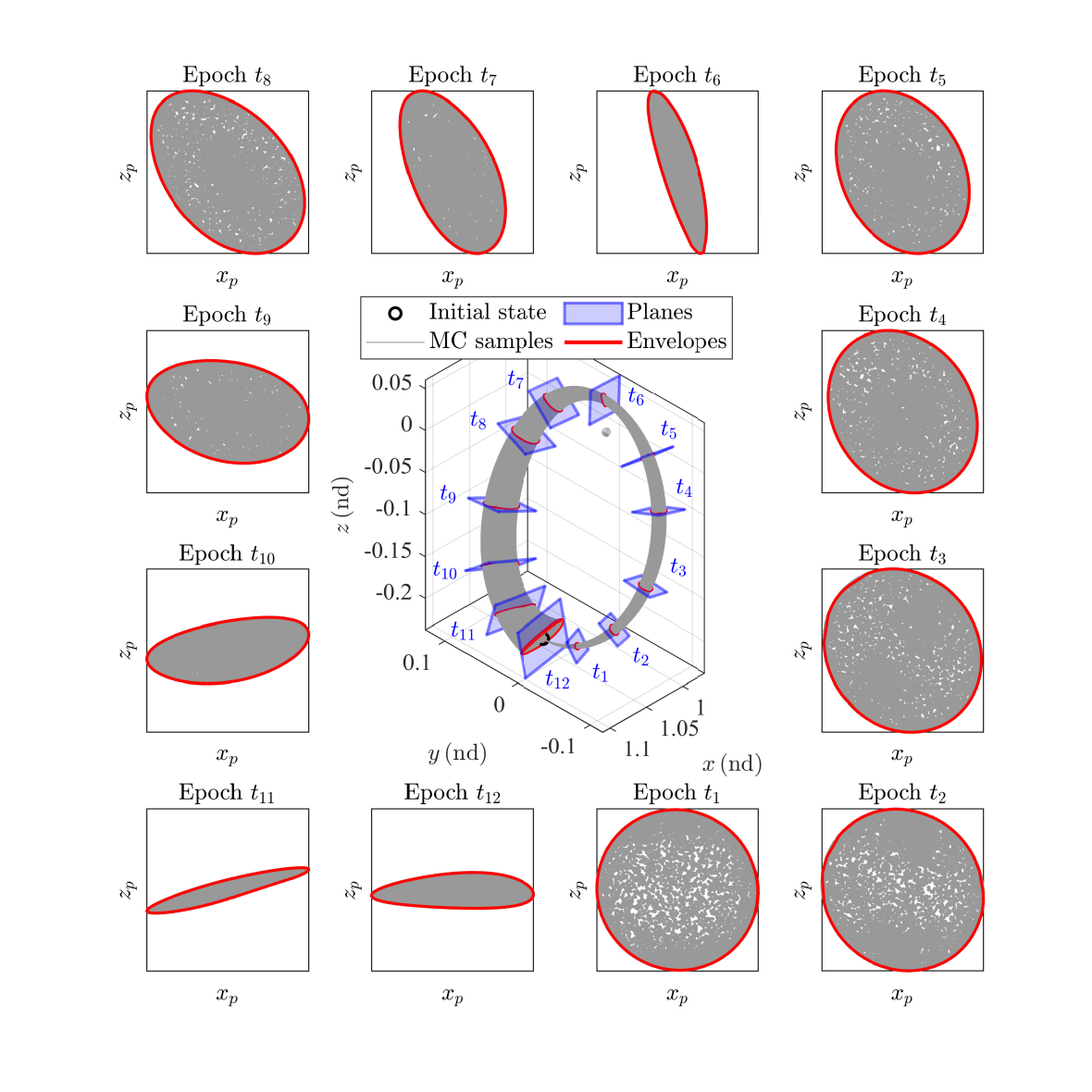}
    \caption{\label{fig:Characterized planes and the corresponding RS’s envelopes at different epochs} Auxiliary planes and the corresponding RS’s envelopes at twelve different epochs.}
\end{figure}

\begin{figure}[!h]
    \centering
    \includegraphics[width=0.48\linewidth]{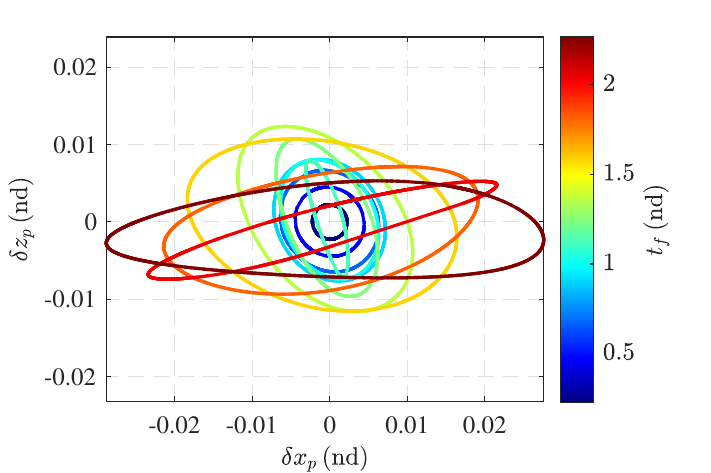}
    \caption{\label{fig:A comparison of the RS’s envelopes at different epochs} A comparison of the RS’s envelopes at different epochs.}
\end{figure}

To analyze the evolution of the RS, the linearly stable NRHO’s period is discretized into 100 equal-spaced epochs, and for each epoch, the RS envelope is determined using the proposed method. When the epoch is larger than $0.15T$, the threshold is set as $\epsilon=10^{-5}$; otherwise, $\epsilon=10^{-6}$. This is because when the propagated arc is short, the RS is small; thus, a threshold of $\epsilon=10^{-5}$ is a little large compared with the size of RS. The effects of the threshold $\epsilon$ will be discussed later. Figure~\ref{fig:Time history of the number of sub-domains of the state case} shows the number of sub-domains at different epochs. For cases with the same threshold, more sub-domains are generated when the time span is longer. The RS envelopes of the 100 discrete epochs are shown in Figu.~\ref{fig:Envelopes of the state RSs at different epochs}. Figure~\ref{fig:Characterized planes and the corresponding RS’s envelopes at different epochs} details auxiliary planes, RS envelopes, and 10,000 MC cloud points at twelve epochs $t_k$ ($k \in \{ 1,2, \cdots ,12\} $). Moreover, the RS envelopes at the twelve epochs are plotted in a single figure for comparison. As shown in Fig.~\ref{fig:A comparison of the RS’s envelopes at different epochs}, the shape of the RS envelope is approximately circular at the beginning, and it stretches as the time span evolves. 

\begin{figure}[!h]
    \centering
    \includegraphics[width=0.9\linewidth]{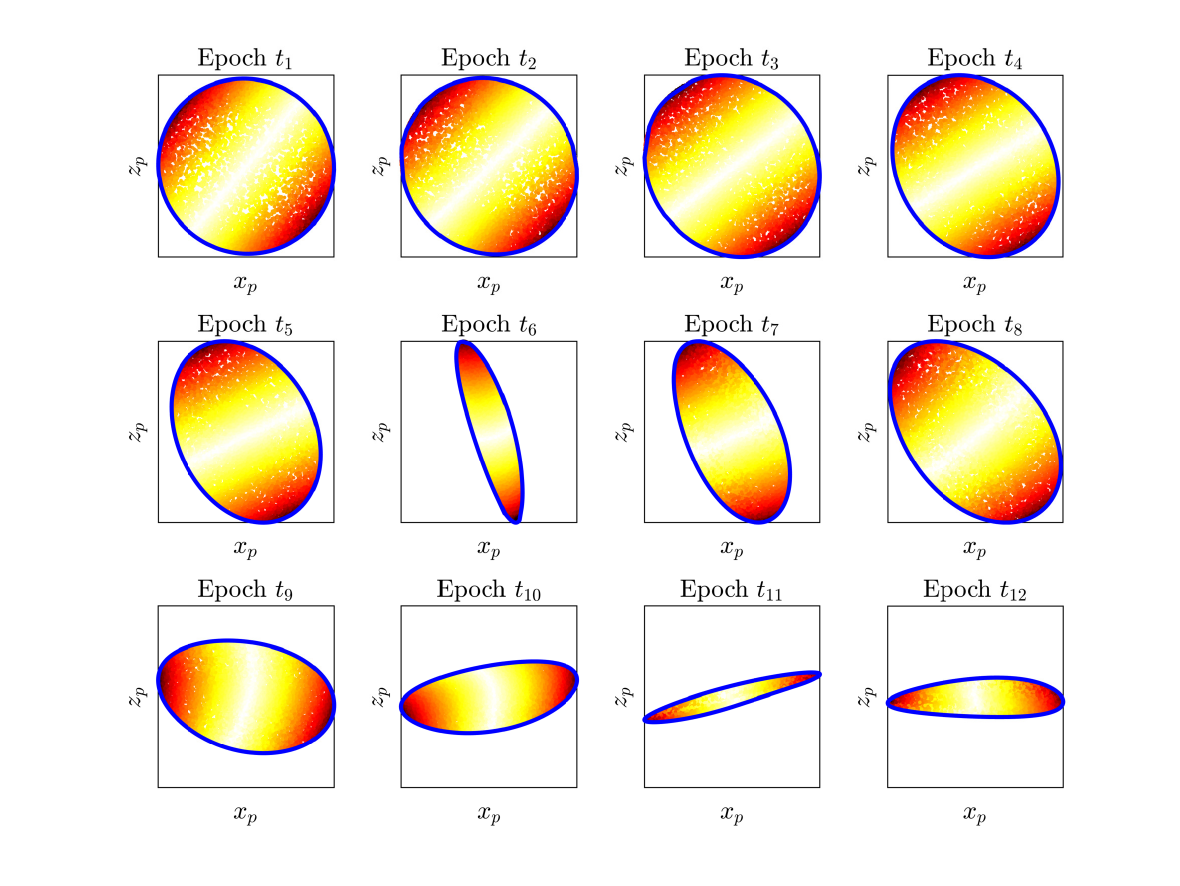}
    \caption{\label{fig:Effects of sensitivity directions on the stretching of the RS envelopes} Effects of sensitivity directions on the stretching of the RS envelopes.}
\end{figure}

Additional analysis is performed to understand the effects of the impulse directions on the RS envelope. Using the Cauchy-Green tensor (CGT), the most sensitive direction, ${\boldsymbol{d}_s} \in {\mathbb{R}^3}$, is determined. Note that the opposite direction of ${\boldsymbol{d}_s}$ (\emph{i.e.}, ${-\boldsymbol{d}_s}$) is also considered the most sensitive direction. This is because the CGT linearizes the dynamics to identify sensitivity. For each MC point in Fig.~\ref{fig:Characterized planes and the corresponding RS’s envelopes at different epochs}, the separation angle between the direction of its velocity impulse and the most sensitive direction is computed, expressed as
\begin{equation} \label{eq:separation_angle}
    \min \left\{ {{{\cos }^{ - 1}}({\boldsymbol{d}_s} \cdot {\boldsymbol{d}_i}),{{\cos }^{ - 1}}( - {\boldsymbol{d}_s} \cdot {\boldsymbol{d}_i})} \right\} \,,
\end{equation}
where ${\boldsymbol{d}_i} \in {\mathbb{R}^3}$ represents the direction of the impulsive maneuver of \emph{i}-th MC point. As shown in Fig.~\ref{fig:Effects of sensitivity directions on the stretching of the RS envelopes}, the MC cloud points are colored, where darker colors represent a smaller separation angle and lighter colors represent a less sensitive direction. One can see from Fig.~\ref{fig:Effects of sensitivity directions on the stretching of the RS envelopes} that the stretching of the RS envelopes is mainly due to the velocity impulse along (or near) the sensitive directions.

\begin{figure}[!h]
    \centering
    \includegraphics[width=0.48\linewidth]{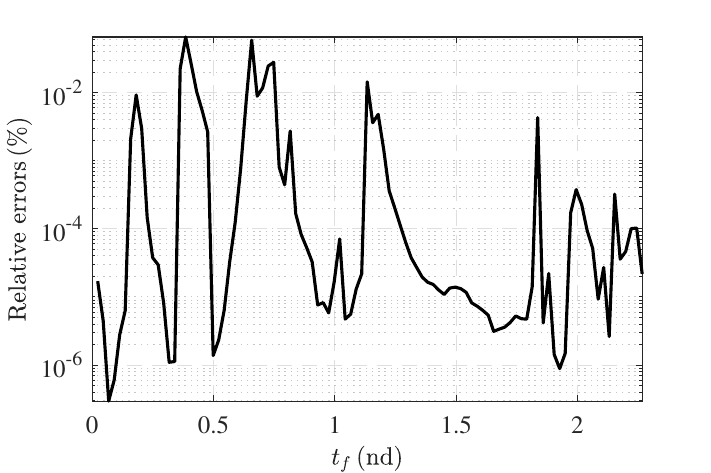}
    \caption{\label{fig:Time histories of the relative errors} Time histories of the relative errors.}
\end{figure}

Figure~\ref{fig:Time histories of the relative errors} presents the profile of the relative errors (\emph{i.e.}, Eq.~\eqref{eq:error_index}), and the values of the error indexes of the twelve epochs are detailed in Table~\ref{tab:Error indexes at twelve different epochs}. The maximal relative error is 0.0658\%, and the average relative error is 0.0032\%. In 91\% of cases, the relative error index $P$ is smaller than 0.01\%. As discussed before, the proposed method cannot provide exact RS envelopes; instead, it approximates RS envelopes as the high-order polynomials will lose some accuracy.

\begin{table}[!h]
    \caption{\label{tab:Error indexes at twelve different epochs} Error indexes at twelve different epochs}
    \centering
    \begin{tabular}{lccc}
        \hline\hline
        Epoch & $S_{\rm{RS}} \,\rm{(nd^{2})}$ & $d_{\max} \,(\rm{nd})$ & $P$ (\%) \\ \hline
        $t_{1}$ & 1.5595$\times 10^{-5}$ & 4.7827$\times 10^{-6}$ & 1.4667$\times 10^{-4}$ \\
        $t_{2}$ & 6.2254$\times 10^{-5}$ & 5.8257$\times 10^{-5}$ & 5.4517$\times 10^{-3}$ \\
        $t_{3}$ & 1.2948$\times 10^{-4}$ & 1.0729$\times 10^{-4}$ & 8.8901$\times 10^{-3}$ \\
        $t_{4}$ & 1.7466$\times 10^{-4}$ & 7.5522$\times 10^{-6}$ & 3.2656$\times 10^{-5}$ \\
        $t_{5}$ & 1.4828$\times 10^{-4}$ & 1.0267$\times 10^{-5}$ & 7.1087$\times 10^{-5}$ \\
        $t_{6}$ & 4.0509$\times 10^{-5}$ & 7.6442$\times 10^{-5}$ & 1.4425$\times 10^{-2}$ \\
        $t_{7}$ & 1.7955$\times 10^{-4}$ & 1.9142$\times 10^{-5}$ & 2.0407$\times 10^{-4}$ \\
        $t_{8}$ & 3.8746$\times 10^{-4}$ & 8.7055$\times 10^{-6}$ & 1.9560$\times 10^{-5}$ \\
        $t_{9}$ & 5.9457$\times 10^{-4}$ & 6.5921$\times 10^{-6}$ & 7.3088$\times 10^{-6}$ \\
        $t_{10}$ & 4.8261$\times 10^{-4}$ & 8.2400$\times 10^{-6}$ & 1.4069$\times 10^{-5}$ \\
        $t_{11}$ & 1.8523$\times 10^{-4}$ & 1.3197$\times 10^{-5}$ & 9.4025$\times 10^{-5}$ \\
        $t_{12}$ & 5.4043$\times 10^{-4}$ & 1.0832$\times 10^{-5}$ & 2.1712$\times 10^{-5}$ \\
        \hline\hline
    \end{tabular}
\end{table}

\begin{figure}[!h]
    \centering
    \includegraphics[width=0.48\linewidth]{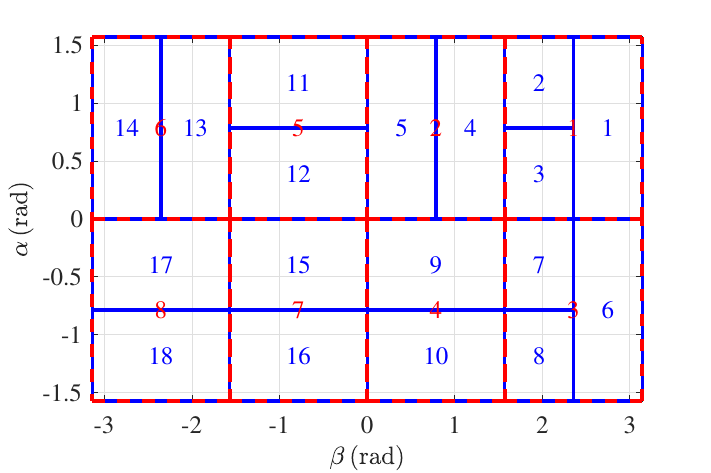}
    \caption{\label{fig:Sub-domains of the short-term propagation case under different thresholds} Sub-domains of the short-term propagation case (red: $\epsilon=10^{-5}$; blue: $\epsilon=10^{-6}$).}
\end{figure}

\begin{figure}[!h]
    \centering
    \subfigure[Final envelopes]{
        \label{fig:22a}
        \includegraphics[width=0.48\linewidth]{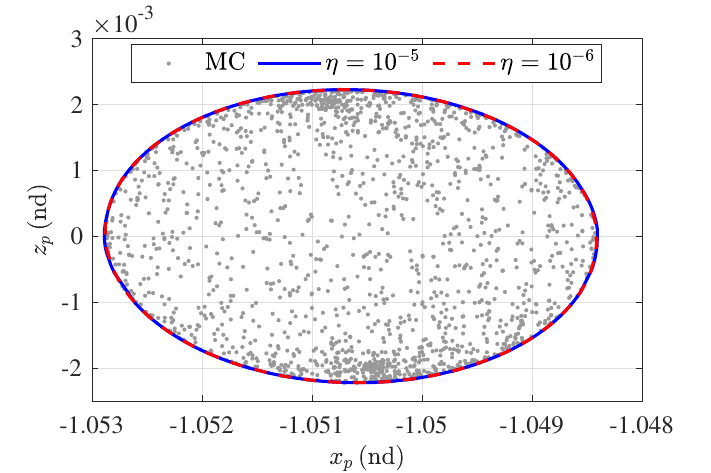}
    }
    \subfigure[Larger plot of \subref{fig:22a}]{
        \label{fig:22b}
        \includegraphics[width=0.48\linewidth]{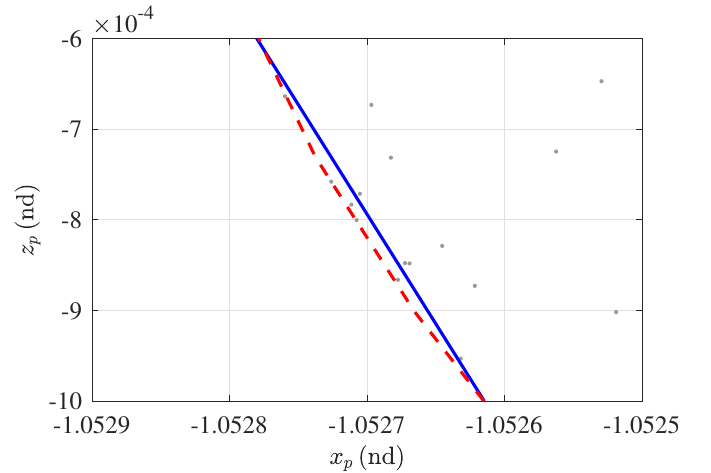}
    }
    \caption{\label{fig:Envelopes of the short-term propagation case under different thresholds} Envelopes of the short-term propagation case under different thresholds.}
\end{figure}

The relative errors can be reduced by setting a smaller threshold or increasing the maximal order of the polynomials. A short-term propagation case with $t_f=0.1T$ is discussed to show the effects of the threshold on the accuracy of the determined RS. Two thresholds, $\epsilon=10^{-5}$ and $\epsilon=10^{-6}$, are simulated. The split sub-domains under different thresholds are illustrated in Fig.~\ref{fig:Sub-domains of the short-term propagation case under different thresholds}. Decreasing the threshold by one order of magnitude, the number of sub-domains increases from 8 (blue boxes in Fig.~\ref{fig:Sub-domains of the short-term propagation case under different thresholds}) to 18 (red boxes in Fig.~\ref{fig:Sub-domains of the short-term propagation case under different thresholds}). The final envelopes under two thresholds are compared in Fig.~\ref{fig:Envelopes of the short-term propagation case under different thresholds}, and the corresponding relative error results are listed in Table~\ref{tab:Error indexes of the short-term propagation case under different threshold}. It can be seen from Fig.~\ref{fig:Envelopes of the short-term propagation case under different thresholds}\subref{fig:22b} that the red envelope ($\epsilon=10^{-6}$) wraps more MC points than the blue envelope ($\epsilon=10^{-5}$). Additionally, Table~\ref{tab:Error indexes of the short-term propagation case under different threshold} shows that the relative error decreases by 76.14\% (from $6.8445\time 10^{-4}$ to $1.6328\time 10^{-4}$).

\begin{table}[!h]
    \caption{\label{tab:Error indexes of the short-term propagation case under different threshold} Error indexes of the short-term propagation case under different threshold}
    \centering
    \begin{tabular}{lccc}
        \hline\hline
        Threshold & $S_{\rm{RS}} \,\rm{(nd^{2})}$ & $d_{\max} \,(\rm{nd})$ & $P$ (\%) \\ \hline
        $\epsilon=10^{-5}$ & 1.5606$\times 10^{-5}$ & 1.0335$\times 10^{-5}$ & 6.8445$\times 10^{-4}$ \\
        $\epsilon=10^{-6}$ & 1.5595$\times 10^{-5}$ & 5.0463$\times 10^{-6}$ & 1.6328$\times 10^{-4}$ \\
        \hline\hline
    \end{tabular}
\end{table}

\subsection{Reachable Set in State Space: 9:2 NRHO Case}
\label{Sec:Reachable Set in State Space: 9:2 NRHO Case}
\begin{figure}[!h]
    \centering
    \includegraphics[width=0.48\linewidth]{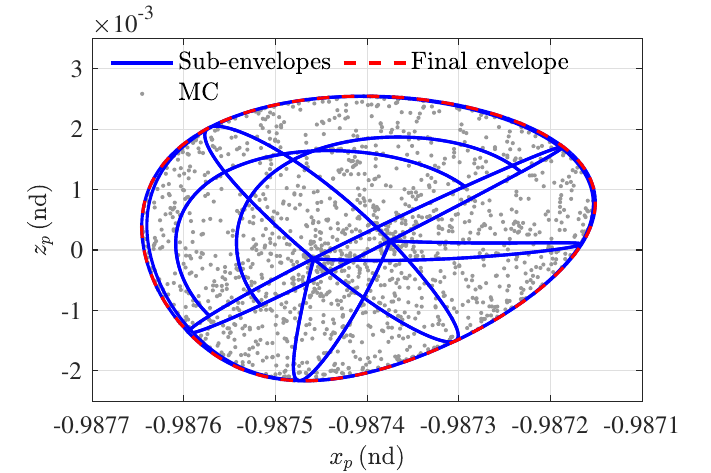}
    \caption{\label{fig:Final envelope of the 9:2 NRHO at its perlune} Final envelope of the 9:2 NRHO at its perlune.}
\end{figure}

\begin{figure}[!h]
    \centering
    \includegraphics[width=0.48\linewidth]{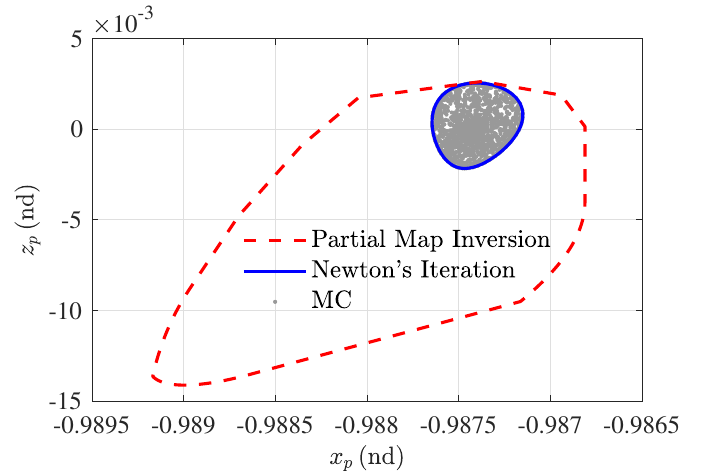}
    \caption{\label{fig:Comparison between the partial map inversion and Newton’s iteration approaches} Comparison between the partial map inversion and Newton’s iteration approaches.}
\end{figure}

In this subsection, a 9:2 NRHO is taken as the nominal orbit, with its initial orbital states given in Table~\ref{tab:Initial orbital state of the 9:2 NRHO}. The propagated time span is set as half of the orbital period, say, $t_f=0.5T$. Thus, at the final epoch $t_f$, the 9:2 NRHO reaches its perilune, where dynamics are highly nonlinear due to the strong gravitational accelerations from the Moon. The threshold for the ADS is set as $\eta =10^{-5}$, and the Newton’s iteration approach is employed to derive polynomials. The envelope of the RS at $t_f=0.5T$ is shown in Fig.~\ref{fig:Final envelope of the 9:2 NRHO at its perlune}. Additionally, the RS envelopes based on the two modeling approaches (\emph{i.e.}, the partial map inversion and Newton’s iteration approaches) are compared in Fig.~\ref{fig:Comparison between the partial map inversion and Newton’s iteration approaches}. One can see from Fig.~\ref{fig:Comparison between the partial map inversion and Newton’s iteration approaches} that after implementing partial map inversion, the RS is overestimated, indicating that the Newton’s iteration approach outweighs the map inversion approach in terms of accuracy when handling RS problems in highly nonlinear dynamics.

\subsection{Reachable Set in Observation Space}
\label{Sec:Reachable Set in Observation Space (simulation)}
\begin{figure}[!h]
    \centering
    \includegraphics[width=0.48\linewidth]{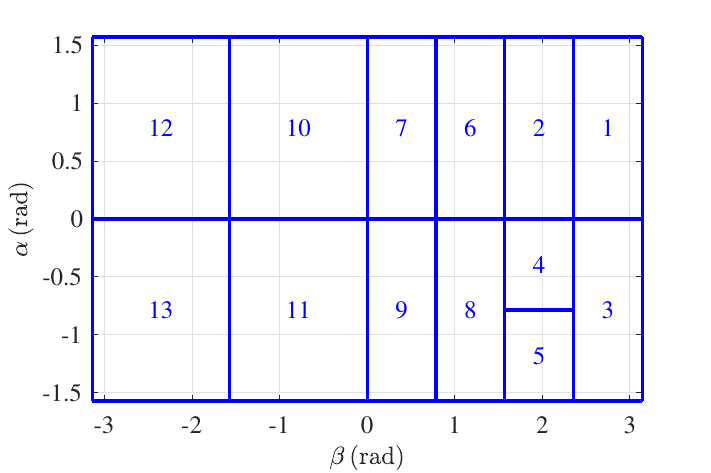}
    \caption{\label{fig:Split Sub-domains of the observation case} Split Sub-domains of the observation case.}
\end{figure}

\begin{figure}[!h]
    \centering
    \subfigure[Initial guesses, anchor points, and approximated characteristic points]{
        \label{fig:24a}
        \includegraphics[width=0.48\linewidth]{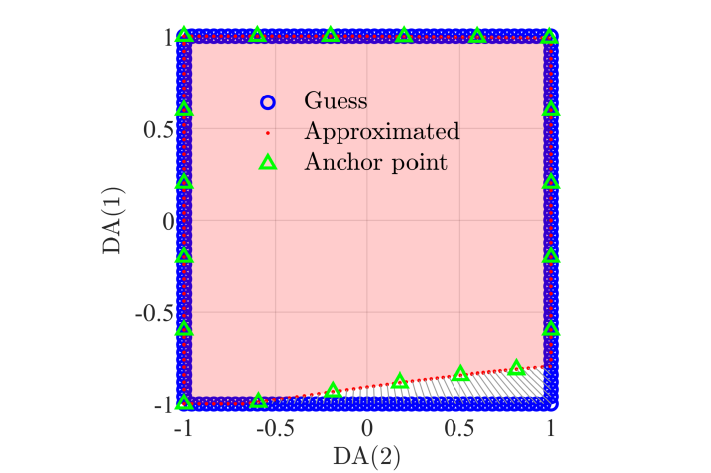}
    }
    \subfigure[Approximated and solved (exact) characteristic points]{
        \label{fig:24b}
        \includegraphics[width=0.48\linewidth]{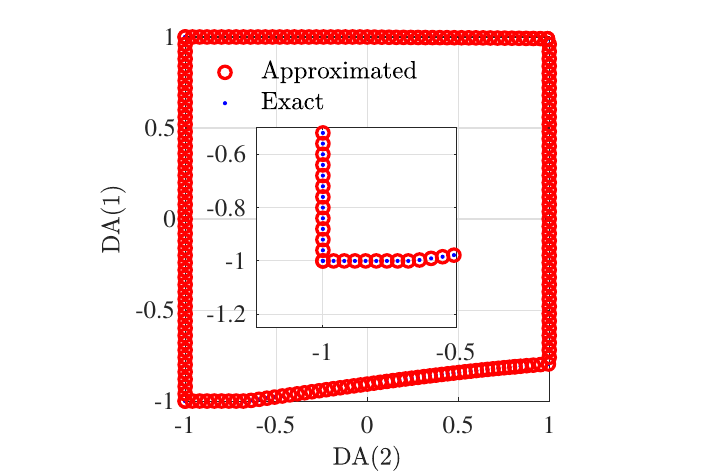}
    }
    \caption{\label{fig:Characteristic curve of the 1st sub-domain} Characteristic curve of the $1^{\rm{st}}$ sub-domain.}
\end{figure}

\begin{figure}[!h]
    \centering
    \subfigure[Sub-envelope]{
        \label{fig:25a}
        \includegraphics[width=0.48\linewidth]{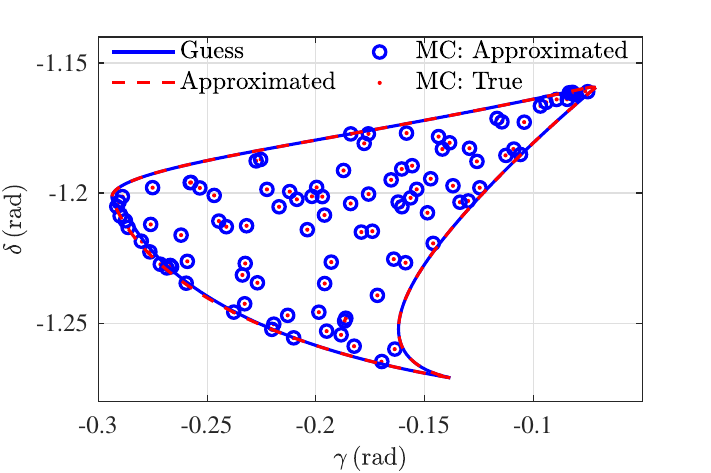}
    }
    \subfigure[Larger plot of \subref{fig:25a}]{
        \label{fig:25b}
        \includegraphics[width=0.48\linewidth]{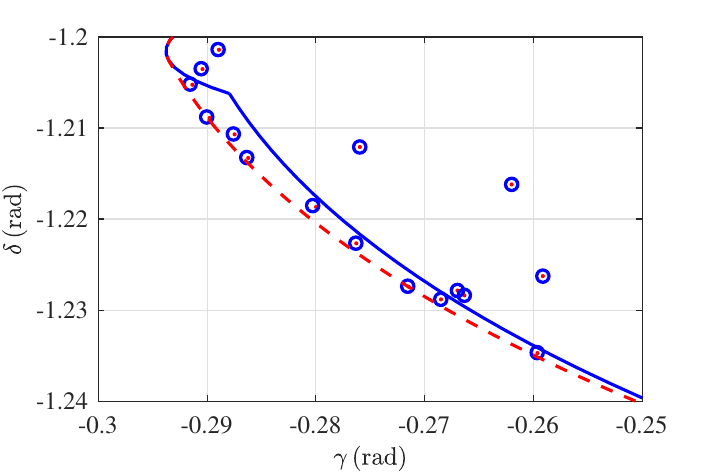}
    }
    \caption{\label{fig:Sub-envelope of the 1st sub-domain} Sub-envelope of the $1^{\rm{st}}$ sub-domain.}
\end{figure}

Using the parameters in Table~\ref{tab:Initial orbital state of the linearly stable NRHO}, Table~\ref{tab:Initial orbital state of the 9:2 NRHO}, and Table~\ref{tab:User-defined parameters for simulations}, and following the procedures in Table~\ref{tab:Pseudocode to determine the boundary of RS in state space}, the RS in the (angle) observation space at the epoch ${t_f} = 1T$ is determined and presented in this subsection. Figure~\ref{fig:Split Sub-domains of the observation case} shows the sizes of 13 split sub-domains with the threshold set as $\epsilon=10^{-3}$. Figure~\ref{fig:Characteristic curve of the 1st sub-domain} and Fig.~\ref{fig:Sub-envelope of the 1st sub-domain} give the characteristic curve and sub-envelope of the first sub-domain, respectively. One can see from Fig.~\ref{fig:Characteristic curve of the 1st sub-domain} that the approximated and solved characteristic points match well. In addition, as shown in Fig.~\ref{fig:Sub-envelope of the 1st sub-domain}, the MC points can be accurately approximated by the high-order polynomials and are included in the identified sub-envelope.

\begin{figure}[!h]
    \centering
    \includegraphics[width=0.48\linewidth]{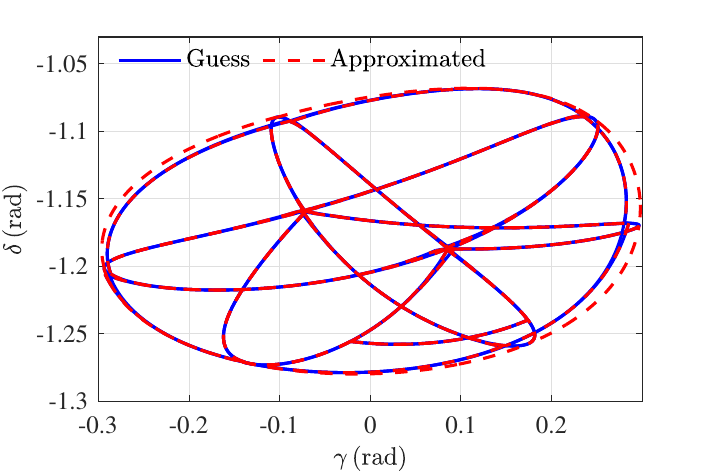}
    \caption{\label{fig:Sub-envelopes of all sub-domains of the observation case} Sub-envelopes of all sub-domains of the observation case.}
\end{figure}

\begin{figure}[!h]
    \centering
    \includegraphics[width=0.48\linewidth]{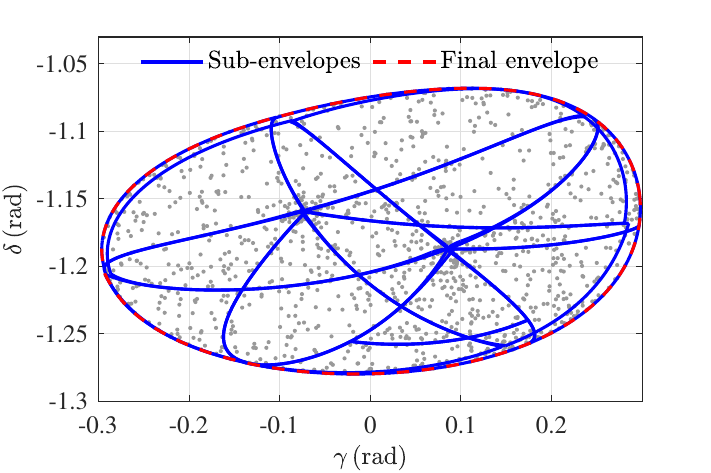}
    \caption{\label{fig:Final envelope of the observation case} Final envelope of the observation case.}
\end{figure}

\begin{figure}[!h]
    \centering
    \includegraphics[width=0.48\linewidth]{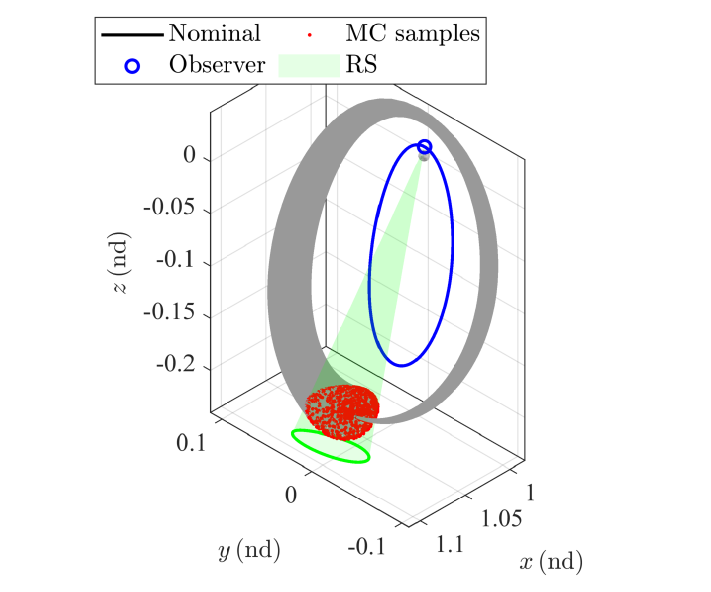}
    \caption{\label{fig:3D plot of the RS in observation space} 3D plot of the RS in observation space.}
\end{figure}

After determining the sub-envelopes of all sub-domains (as shown in Fig.~\ref{fig:Sub-envelopes of all sub-domains of the observation case}), the final envelope is obtained (as shown in Fig.~\ref{fig:Final envelope of the observation case}). Apart from Fig.~\ref{fig:Final envelope of the observation case}, which illustrates the RS using the azimuth and elevation, Fig.~\ref{fig:3D plot of the RS in observation space} shows the observation RS from a 3D view (\emph{i.e.}, the green cone). Note that in Fig.~\ref{fig:Final envelope of the observation case} and Fig.~\ref{fig:3D plot of the RS in observation space}, 1,300 MC points are presented (still 100 points for each sub-domain), and based on these MC points, the relative error of the observation RS envelope is $4.8443 \time 10^{-7}$ ($S_{\rm{RS}}=0.0986\,\rm{rad}^2$ and $d_{\max}=2.1857 \times 10^{-4}\, \rm{rad}$).

\section{Conclusion}
\label{Sec:Conclusion}
A method has been proposed for determining the single-impulse reachable set (RS) in both state and observation spaces under arbitrary dynamics. This approach utilizes differential algebra and automatic domain splitting techniques to derive high-order Taylor polynomials for the targeted variables with respect to impulsive maneuver parameters. It then employs polynomial pieces to approximate the solutions of the envelope equation, thereby identifying the boundaries of the RS. The method has been effectively applied to calculate the envelope of the single-impulse RS for two near-rectilinear halo orbits (NRHOs) in Cislunar space. Simulation results indicate that the RS of the NRHO can be accurately approximated, with relative errors not exceeding 0.0658\%. Furthermore, these relative errors can be reduced by approximately 76.14\% by applying a smaller threshold within the method. In comparison to directly solving the envelope equation, the use of polynomial pieces can reduce computational costs by more than 84\%. The partial map inversion approach can effectively approximate the motion of the linearly stable NRHO; however, it breaks at the perilune of the 9:2 NRHO, where Newton’s iteration method works. Additionally, the RS envelopes stretch as the time span increases, primarily due to the velocity impulse along sensitive directions.

\section*{Acknowledgments}
This work was supported by the National Natural Science Foundation of China (No. 12150008, No. 62394353, No. 124B2049), the Changjiang Scholars Program (No. T2023191) and the Beijing Institute of Technology Research Fund Program for Innovative Talents (No. 2022CX01008). Xingyu Zhou is grateful for the financial support provided by the China Scholarship Council (No. 202406030186) and thanks Xiaoyu Fu for his helpful discussions.

\bibliography{sample}

\end{document}